\documentclass[12pt]{article}
\newcommand{\BE}{\begin{eqnarray}}
\newcommand{\EE}{\end{eqnarray}}
\newcommand{\nn}{\nonumber}
\newcommand{\ps}{\partial_{\sigma}}
\newcommand{\pt}{\partial_{\tau}}
\usepackage{amssymb}
\usepackage{amsmath}
\usepackage{array}
\usepackage{latexsym,graphicx,epsfig}
\usepackage{hyperref}
\usepackage{subfigure}
\usepackage[sort]{cite}

\begin{document}
\vspace*{-.6in} \thispagestyle{empty}
\begin{flushright}
UK-06-05
\end{flushright}
\baselineskip = 18pt

\vspace{1.0in} {\Large
\begin{center}
Kinky Strings in $AdS_5\times S^5$
\end{center}} \vspace{.5in}

\begin{center}
Tristan McLoughlin$^1$, and Xinkai Wu$^2$
\\
\emph{$^1$Department of Physics\\Pennsylvania State University, University Park, PA 16802, USA\\
$^2$Department of Physics and Astronomy\\
University of Kentucky, Lexington, KY  40506, USA\\
tmclough@phys.psu.edu,
xinkaiwu@pa.uky.edu}
\end{center}
\vspace{0.4in}

\begin{center}
\textbf{Abstract}
\end{center}
\begin{quotation}
\noindent We construct a family of closed string solutions with
kinks in a subspace of $AdS_5\times S^5$ and study their
properties. In certain limits these solutions become folded
pulsating strings, although in general they are made of multiple pulsating rectangles. 
One unusual feature of these solutions is that 
their monodromy matrices are trivial, leading to
vanishing quasi-momenta. Exact B\"{a}cklund transformations of these
solutions are found, again giving vanishing higher conserved
charges. We also consider
the fluctuation modes around these solutions as well as the
semiclassical splitting of these strings.
\end{quotation}

\newpage

\pagenumbering{arabic}

%%%%%%%%%%%%%%%%%%%%%%%%%%%%%%%%%%%%%%%%%%%%

\section{Introduction}
In $D$-dimensional flat space the general solution to the bosonic
string equations of motion is specified by two unit modulus vectors
${\bf a}^{(1)}$ and ${\bf b}^{(1)}$, which are the first derivatives
w.r.t. the worldsheet left moving and right moving coordinates.
Being first derivatives, ${\bf a}^{(1)}$ and ${\bf b}^{(1)}$ do not
have to be continuous. When they have discontinuities, the string
profile is still continuous, although now with kinks. A kink is the
joining point of two string segments where the angle between the two
segments is nonzero. People have also considered strings with spikes
(sometimes also called cusps), with the spikes being the joining
points of two anti-parallel string segments. Both kinky strings and
spiky strings in flat space have been investigated quite extensively
in the cosmic string context, see for example the review
\cite{Hindmarsh:1994re}. One important difference between kinky
strings and spiky strings is that, while the vectors ${\bf a}^{(1)}$
and ${\bf b}^{(1)}$ are discontinuous functions of worldsheet
coordinates for the kinky strings, they are continuous functions for
the spiky strings (spikes develop when ${\bf a}^{(1)}$ and
$-{\bf b}^{(1)}$, which ``rotate'' with different frequencies, cross
each other). We shall see this in more detail in Section
\ref{section_flatspace}.
In \cite{Kruczenski:2004wg,Ryang:2005yd} the spiky strings in flat space were
generalized to $AdS_5\times S^5$ and their role in the AdS/CFT
correspondence was investigated. It is natural to ask whether
similar things can be done for the kinky strings, which provides 
motivation for the current paper. We shall generalize the kinky
strings to the $R_t\times S^5$ sector of $AdS_5\times S^5$ and study
their properties such as higher conserved charges, stability under
fluctuations, and semi-classical splitting. Of course the study of semi-classical
strings in $AdS_5\times S^5$ is of particular interest as it may shed light on the
$AdS/CFT$ duality away from the BPS limit and there has been a great deal of progress
in understanding semi-classical strings with 
large quantum numbers such as angular momentum or spin \cite{Gubser:2002tv,Frolov:2002av, 
Kruczenski:2004wg,Frolov:2003qc,Beisert:2003xu,Kruczenski:2003gt,Tseytlin:2003ii,
Minahan:2002rc} (see \cite{Tseytlin:2004xa} for a review and a more complete set of 
references). Those semi-classical strings were mapped to dual ``long'' gauge theory operators 
and string energies matched to the operator anomalous dimensions. It would be interesting 
to study the operators dual to the kinky strings however it is at present unclear to us  
what these operators should be and so this side of the duality will be left to a
future work.

One of the most remarkable facets of the ${\cal N}=4$ --- $AdS_5\times S^5$ duality is that 
both sides appear to be integrable with the gauge theory being described by an integrable
spin chain \cite{Minahan:2002ve,Beisert:2003yb}. The string theory has been shown to be classically integrable \cite{Bena:2003wd,Mandal:2002fs} and there exists some evidence that this persists
quantum mechanically \cite{Berkovits:2004xu}. One consequence of integrability is the
existence of an infinite number of conserved charges and in general there are several 
methods to construct these charges. In particular we will concern ourselves with the
Lax method involving the monodromy matrix, which has previously been considered for the
string theory in \cite{Kazakov:2004qf,Kazakov:2004nh, Beisert:2004ag,Schafer-Nameki:2004ik,Beisert:2005bm}, and the 
method involving the B\"{a}cklund transformations 
\cite{Pohlmeyer:1975nb,Arutyunov:2003rg,Arutyunov:2005nk,Mikhailov:2005wn,Mikhailov:2005zd}.
 One peculiar feature of the kinky solutions is that their
monodromy matrices are trivial, leading to vanishing quasi momenta.
We also solve the B\"{a}cklund transformation for these solutions in an
exact manner and show that the associated higher charges all vanish,
which is consistent with the finding of trivial monodromies in light
of the relation between the monodromy and B\"{a}cklund charges explained
in \cite{Arutyunov:2005nk}. This feature is in contrast with the quite general 
constructions made previously in the literature (e.g.
\cite{Kazakov:2004qf},\cite{Kazakov:2004nh, Beisert:2004ag,Schafer-Nameki:2004ik, Beisert:2005bm}) 
where the quasi momentum has a pole as the spectral parameter approaches to $\pm 1$ and
is certainly not vanishing. The
reason for this apparent conflict is that
the string embedding can have discontinuities in its first
worldsheet derivatives, i.e. the string profile can have kinks.
Section \ref{subsection_monodromy} explains this issue in detail.
Semi-classical analysis of string decay in both flat spacetime and
$AdS_5\times S^5$ has been performed in \cite{Iengo:2002tf}, \cite{Iengo:2003ct}, \cite{Chialva:2003hg}, 
\cite{Chialva:2004xm}, 
\cite{Iengo:2006gm}, and \cite{Peeters:2004pt}, and we carry out
this analysis for the splitting of the kinky strings.

This paper is organized as follows. In Section
\ref{section_flatspace} we review classical bosonic
 string solutions with spikes and those with kinks in flat spacetime. In Section
\ref{section_R_times_S5} we then generalize these kinky solutions to the
$R_t\times S^5$ sector of $AdS_5\times S^5$. More specifically, we
construct a class of solutions with kinks to the equations of
motion, find their B\"{a}cklund transformations as well as their monodromy matrices, and also give some explicit
example solutions. In Section \ref{section_NR_system} we compare our
solutions with the pulsating strings obtained via the ``2d-dual''
version of Neumann-Rosochatius(NR) system. In Section
\ref{section_fluctuation} we investigate the solutions' stability
under small fluctuations. In Section \ref{section_decay_of_strings}
we carry out a semi-classical analysis of the splitting of these
strings. We end with Section \ref{section_discussion} discussing the connection of the kinky strings to the $2d$ flat space strings 
with point-like longitudinal degrees of freedom \cite{Patrascioiu:1974un}, \cite{Bardeen:1975gx},\cite{Bardeen:1976yt}, 
\cite{Bars:1993sq}, boost of the kinky solutions to give them angular momenta on $S^5$ and comparison
to the ``giant magnons'' constructed in \cite{Hofman:2006xt}, as well as other interesting flat space
solutions with kinks whose $AdS_5\times S^5$ generalizations are still to be found.

\section{String solutions in flat space}\label{section_flatspace}

In this section we review classical string solutions in flat space following mainly
the review article \cite{Hindmarsh:1994re}. The target space is taken to be
$D$-dimensional Minkowski 
\BE
ds^2=\eta_{\mu\nu}dX^\mu dX^\nu;\ \ \eta_{\mu\nu}=(-1,1,...,1)
\EE
with $\mu,\nu=0,1,...,D$. The Nambu-Goto action of the string is
\BE
S=-T\int d\tau d\sigma\sqrt{-\det h_{\alpha\beta}}\ ;\ \ h_{\alpha\beta}=\eta_{\mu\nu}
\partial_\alpha X^\mu \partial_\beta X^\nu
\EE
with $\alpha,\beta=0,1$ being worldsheet indices. We use the conformal gauge
\BE
& &\eta_{\mu\nu}(\partial_\tau X^\mu\partial_\tau X^\nu+\partial_\sigma X^\mu\partial_\sigma X^\nu)=0\nonumber\\
& &\eta_{\mu\nu}\partial_\tau X^\mu\partial_\sigma X^\nu=0
\EE
which as usual is written as
\BE\label{eqn_conformal_gauge}
& &\dot{X}^2+X'^2=0\nonumber\\
& &\dot{X}\cdot X'=0
\EE
and in the conformal gauge the equation of motion is
\BE\label{eqn_eom_flat}
\ddot{X}^\mu-X''^{\mu}=0
\EE
The conformal gauge still leaves some freedom which we further restrict by setting $t=\tau$ (note $t$ is just $X^0$).
Then the gauge condition (\ref{eqn_conformal_gauge}) becomes
\BE\label{eqn_new_conformal_gauge}
\dot{\bf X}^2+{\bf X}'^2=1,\ \ \dot{\bf X}\cdot{\bf X}'=0
\EE
while equation (\ref{eqn_eom_flat}) now reads
\BE\label{eqn_new_eom}
\ddot{\bf X}-{\bf X}''=0
\EE
where we have used $\bf X$ to denote the $(D-1)$-component spatial vector. As can be seen, the general solution
\BE\label{eqn_general_solution}
{\bf X}=\frac{1}{2}\left[{\bf a}(u)+{\bf b}(v)\right],
\EE
with $u=\sigma-\tau, v=\sigma+\tau$, satisfies the e.o.m. for arbitrary ${\bf a},{\bf b}$. The gauge condition (\ref{eqn_new_conformal_gauge})
imposes the constraint that
\BE\label{eqn_unit_constraint_ab}
\left({\bf a}^{(1)}\right)^2=1,\ \ \left({\bf b}^{(1)}\right)^2=1
\EE
where the superscript $(1)$ means first derivative with respect to the argument. We shall consider a closed string with
$\sigma$ going from $0$ to $L$. Then the periodic boundary condition $\int_0^L d\sigma {\bf X}'=0$ (we do not compactify
the target space; hence there is no winding) together with our choice to work in the string's center of mass frame
$\int_0^L d\sigma\dot{\bf X}=0$ implies that
\BE\label{eqn_center_constraint_ab}
\int_0^L d\sigma\ {\bf a}^{(1)}=0,\ \ \int_0^L d\sigma\ {\bf b}^{(1)}=0
\EE
Pictorially the constraints (\ref{eqn_unit_constraint_ab}) and (\ref{eqn_center_constraint_ab}) mean that ${\bf a}^{(1)}(u)$
and ${\bf b}^{(1)}(v)$ are two curves on the sphere ${\bf S}^{D-2}$ with their centers at the origin. The constraint
(\ref{eqn_center_constraint_ab}) must hold at any time $\tau$, which implies that
${\bf a}^{(1)}, {\bf b}^{(1)}$ (and hence ${\bf a},{\bf b}$)
are periodic in their arguments with period $L$. Using the periodicity of ${\bf a}, {\bf b}$, we can show that
\BE
{\bf X}(\sigma+\frac{L}{2},\tau+\frac{L}{2})={\bf X}(\sigma,\tau)
\EE
i.e., the string looks the same at $\tau$ and $\tau+\frac{L}{2}$. Next let us give some interesting examples of solutions.

$\bullet$ {The spiky solution}

One can take the following ${\bf a}^{(1)}$ and ${\bf b}^{(1)}$ \cite{Burden:1985md}
\BE\label{eqn_flatspace_spiky_string}
& &{\bf a}^{(1)}(u)=\cos\left(2\pi M\frac{u}{L}\right){\bf e}_1+\sin\left(2\pi M\frac{u}{L}\right){\bf e}_2\nonumber\\
& &{\bf b}^{(1)}(v)=-\cos\left(2\pi N\frac{v}{L}\right){\bf e}_1+\sin\left(2\pi N\frac{v}{L}\right){\bf e}_2
\EE
Pictorially ${\bf a}^{(1)}$ and $-{\bf b}^{(1)}$ rotate with different angular frequencies.
Whenever they cross each other, ${\bf X}'$ vanishes, $\dot{{\bf X}}$ has magnitude of unity, and the point on the string where this happens moves 
at the speed of light. This is how a spike is developed. When $\frac{M}{N}$ is not an integer, the string will be
self-intersecting. When $\frac{N}{M}$ is an integer, more specifically, let us take $M=1, N=n-1$, the string is not
self-intersecting, and $n$ is the number of spikes. The solution given in eqn. (\ref{eqn_flatspace_spiky_string}) has its
spikes pointing outward. If we flip the sign of the first term in ${\bf b}^{(1)}(v)$, the spikes will be pointing inward.
The spiky strings considered in \cite{Kruczenski:2004wg,Ryang:2005yd} are generalizations of these solutions to $AdS_5\times S^5$.
\footnote{Just like their flat space cousins, the spiky strings in $AdS_5\times S^5$ also have continuous first 
derivatives w.r.t. to worldsheet coordinates. This can be easily seen by considering the special case of two-spikes,
which is the familiar spinning folded string.}

$\bullet$ {The kinky solution}

Note that ${\bf a}^{(1)}$ and ${\bf b}^{(1)}$ do not have to be continuous functions of $u$ and $v$. When they have discontinuities,
kinks are developed on the string. Let us take \cite{Garfinkle:1987yw}
\BE\label{eqn_kinky_a} {\bf a}^{(1)}(u)=& &{\bf e}_1\ \ \ \text{when }mL\le u\le (m+\frac{1}{2})L\nonumber\\
& &-{\bf e}_1\ \ \text{when }(m+\frac{1}{2})L\le u\le (m+1)L
\EE with $m\in{\bf Z}$, and
\BE\label{eqn_kinky_b}
{\bf b}^{(1)}(v)=& &{\bf e}_2\ \ \ \text{when }mL\le v\le (m+\frac{1}{2})L\nonumber\\
& &-{\bf e}_2\ \ \text{when }(m+\frac{1}{2})L\le v\le (m+1)L
\EE with $m\in{\bf Z}$. It is easy to see that this choice of ${\bf a}^{(1)}$ and ${\bf b}^{(1)}$ satisfies the constraints
(\ref{eqn_unit_constraint_ab},\ref{eqn_center_constraint_ab}). The resulting solution ${\bf X}$ lies in the $X^1, X^2$
plane and follows periodically the sequence: square $\rightarrow$ doubled line $\rightarrow$ square $\rightarrow$... In
the remainder of this paper, we shall generalize this solution to strings on $R_t\times S^5$.
%See Figure \ref{kinky}. Note that initially the string is a square, with non-zero velocity.
%\begin{figure}[ht]
%\epsfig{figure=kinky.eps,width=7.0in,angle=0}
%\caption{the solution given by eqn.'s~(\ref{eqn_kinky_a},\ref{eqn_kinky_b}) between $\tau=\frac{L}{8}$ and
%$\tau=\frac{5L}{8}$; recall that the string looks the same at $\tau$ and $\tau+\frac{L}{2}$}
%\label{kinky}
%\end{figure}

\section{String solutions on $R_t\times
S^5$}\label{section_R_times_S5}
\subsection{Construction of the solution}
\label{section_our_solution}
We follow the notation of \cite{Frolov:2003qc} in describing $S^5$ as a hypersurface embedded in ${\mathbf R}^6$. The equations of motion and the constraints are
\BE\label{eqn_o6_model}
& &X_AX_A=X_1^2+...+X_6^2=1\nonumber\\
& &\partial_+\partial_-X_A+(\partial_+X_B\partial_-X_B)X_A=0\nonumber\\
& &\partial_+X_A\partial_+X_A=\kappa^2,\ \  \partial_-X_A\partial_-X_A=\kappa^2
\EE
where we have taken $t=\kappa\tau$, and defined $\xi^\pm=\frac{1}{2}(\tau\pm\sigma)$ (and hence $\partial_\pm=\partial_\tau \pm \partial_\sigma$). We define the new fields $Z$, $X$, and $Y$ which are parametrized by 
\BE
Z=X_1+iX_2=\cos\gamma e^{i\phi_1},\ X=X_3+iX_4=\sin\gamma\cos\psi e^{i\phi_2},\ Y=X_5+iX_6=\sin\gamma\sin\psi e^{i\phi_3}\nn\\
\EE
which corresponds to the metric
\BE
ds^2=d\gamma^2+\cos^2\gamma\ d\phi_1^2+\sin^2\gamma(d\psi^2+\cos^2\psi d\phi_2^2+\sin^2\psi d\phi_3^2)
\EE
on $S^5$.

We now restrict our attention to the subsector of $\psi=0,\phi_1=\phi_2=\phi$, which corresponds to
\BE
X_1=\cos\gamma\cos\phi,\ X_2=\cos\gamma\sin\phi,\ X_3=\sin\gamma\cos\phi,\ X_4=\sin\gamma\sin\phi,\ X_5=0,\ X_6=0
\EE
(hence we are looking at an $S^3\subset S^5$) then the metric becomes
\BE
ds^2=d\gamma^2+d\phi^2
\EE
which is simply a flat metric on a torus.

Denoting $\partial_\pm\gamma$ as $\gamma_\pm$ and so on, one finds that the constraints become
\BE
\partial_+X_A\partial_+X_A=\gamma_+^2+\phi_+^2=\kappa^2,\  \partial_-X_A\partial_-X_A=\gamma_-^2+\phi_-^2=\kappa^2
\EE
and using
\BE
\partial_+X_B\partial_-X_B=\gamma_+\gamma_-+\phi_+\phi_-
\EE
one finds the equations of motion become
\BE\label{eqn_eom}
& &\partial_+\partial_-X_1+(\partial_+X_B\partial_-X_B)X_1=(\gamma_+\phi_-+\gamma_-\phi_+)X_4-\gamma_{+-}X_3-\phi_{+-}X_2=0\nonumber\\
& &\partial_+\partial_-X_2+(\partial_+X_B\partial_-X_B)X_2=-(\gamma_+\phi_-+\gamma_-\phi_+)X_3-\gamma_{+-}X_4+\phi_{+-}X_1=0\nonumber\\
& &\partial_+\partial_-X_3+(\partial_+X_B\partial_-X_B)X_3=-(\gamma_+\phi_-+\gamma_-\phi_+)X_2+\gamma_{+-}X_1-\phi_{+-}X_4=0\nonumber\\
& &\partial_+\partial_-X_4+(\partial_+X_B\partial_-X_B)X_4=(\gamma_+\phi_-+\gamma_-\phi_+)X_1+\gamma_{+-}X_2+\phi_{+-}X_3=0
\EE
which are satisfied if we take
\BE\label{eqn_from_eom}
\gamma_{+-}=0,\ \phi_{+-}=0,\ \gamma_+\phi_-+\gamma_-\phi_+=0
\EE
(We note that (\ref{eqn_from_eom}) is the sufficient condition for (\ref{eqn_eom}), but probably not the necessary condition-- 
naively one would expect to get just two equations for $\gamma$ and $\phi$, not three -- 
which means the solutions we shall present below are probably not exhaustive.
\footnote{In particular, by insisting on (\ref{eqn_from_eom}),
 we might be excluding the generalizations of the flat space pulsating polygonal solutions which shall be discussed in
Section \ref{section_discussion}.} )

Setting
\BE\label{eqn_soln1}
\gamma=\Gamma(\xi^+)+\tilde{\Gamma}(\xi^-),\ \phi=\Phi(\xi^+)+\tilde{\Phi}(\xi^-)
\EE
with $\Gamma,\tilde{\Gamma},\Phi,\tilde{\Phi}$ being arbitrary functions satisfies $\gamma_{+-}=0$ and $\phi_{+-}=0$. The equation$\gamma_+\phi_-+\gamma_-\phi_+=0$ is now satisfied if
\BE
\Gamma'(\xi^+)\tilde{\Phi}'(\xi^-)+\tilde{\Gamma}'(\xi^-)\Phi'(\xi^+)=0
\EE (where primes stand for derivatives)
which implies that
\BE\label{eqn_soln2}
\Phi'(\xi^+)=q\Gamma'(\xi^+),\ \ \tilde{\Phi}'(\xi^-)=-q\tilde{\Gamma}'(\xi^-)
\EE
with $q$ being an arbitrary constant and the conformal constraints become
\BE\label{eqn_soln3}
\left[\Gamma'(\xi^+)\right]^2=\frac{\kappa^2}{1+q^2},\ \
 \left[\tilde{\Gamma}'(\xi^-)\right]^2=\frac{\kappa^2}{1+q^2}
\EE
We shall take the range of $\sigma$ to be $[0,L]$. Since we shall only consider solutions without winding,
we have the periodic boundary condition
$\gamma(\tau,\sigma)=\gamma(\tau,\sigma+L)$ and $\phi(\tau,\sigma)=\phi(\tau,\sigma+L)$, which are satisfied if we
take 
 $\Gamma,\tilde{\Gamma},\Phi,\tilde{\Phi}$ to be periodic in their arguments with the period being $\frac{L}{2}$.
 In summary, the solutions are given by (\ref{eqn_soln1}) with the conditions (\ref{eqn_soln2}) and (\ref{eqn_soln3}),
 as well as the periodic boundary condition just mentioned. It is worth pointing out that, due to the periodicity of
 $\Gamma$, $\tilde{\Gamma}$, $\Phi$ and $\tilde{\Phi}$, $\Gamma'$, $\tilde{\Gamma}'$, $\Phi'$ and $\tilde{\Phi}'$
 must integrate to zero over one period. This means
 that $\Gamma'$ and $\tilde{\Gamma}'$ must jump between $\frac{\kappa}{\sqrt{1+q^2}}$ and
 $-\frac{\kappa}{\sqrt{1+q^2}}$, while $\Phi'$ and $\tilde{\Phi}'$ must jump between $\frac{q\kappa}{\sqrt{1+q^2}}$ and
 $-\frac{q\kappa}{\sqrt{1+q^2}}$, i.e. they are discontinuous. As we shall see, this discontinuity exhibited by
  the first derivatives of the embedding map $X$ is a most salient feature of our solutions.

It is easy to show that for the above solution, $\gamma(\tau+\frac{L}{2},\sigma+\frac{L}{2})=\gamma(\tau,\sigma)$ and
$\phi(\tau+\frac{L}{2},\sigma+\frac{L}{2})=\phi(\tau,\sigma)$, i.e. the string looks the same at
$\tau$ and $\tau+\frac{L}{2}$. Besides, two limits we shall often consider are the $q\to 0$ limit, in which the
string becomes folded along the $\gamma$-axis (i.e. $\phi=0$), and the $q\to\infty$ limit, in which the string becomes
folded along the $\phi$-axis (i.e. $\gamma=0$).

Let us compute the charges of the solution. The energy is given by
\BE
E=S_{50}=\sqrt{\lambda}\int_0^{L}\frac{d\sigma}{L}\left(Y_5\dot{Y}_0-Y_0\dot{Y}_5\right)
\EE with $Y_5=\cos t$ and $Y_0=\sin t$ being two of the $AdS_5$ embedding coordinates (the other four $Y_{1,2,3,4}$ all
vanish), and one finds
$E=\sqrt{\lambda}\ \kappa$.
The angular momenta on the $S^5$ are given by
\BE
J_{AB}=\sqrt{\lambda}\int_0^{L}\frac{d\sigma}{L}\left(X_A\dot{X}_B-X_B\dot{X}_A\right)
\EE
One finds
\BE\label{eqn_J_density}
& &X_1\dot{X}_2-X_2\dot{X}_1=\frac{1}{2}(\phi_+ +\phi_-)\cos^2\gamma=q\gamma'\cos^2\gamma\nonumber\\
& &X_1\dot{X}_3-X_3\dot{X}_1=\frac{1}{2}(\gamma_+ +\gamma_-)\cos^2\phi=\frac{1}{q}\phi'\cos^2\phi\nonumber\\
& &X_1\dot{X}_4-X_4\dot{X}_1=\frac{1}{2}\sin\phi\cos\phi(\gamma_+ +\gamma_-)+\frac{1}{2}\sin\gamma\cos\gamma
(\phi_+ +\phi_-)=\frac{1}{q}\phi'\sin\phi\cos\phi+q\gamma'\sin\gamma\cos\gamma \nonumber\\
& &X_2\dot{X}_3-X_3\dot{X}_2=\frac{1}{2}\sin\phi\cos\phi(\gamma_+ +\gamma_-)-\frac{1}{2}\sin\gamma\cos\gamma
(\phi_+ +\phi_-)=\frac{1}{q}\phi'\sin\phi\cos\phi-q\gamma'\sin\gamma\cos\gamma \nonumber\\
& &X_2\dot{X}_4-X_4\dot{X}_2=\frac{1}{2}\sin^2\phi(\gamma_+ +\gamma_-)=\frac{1}{q}\phi'\sin^2\phi\nonumber\\
& &X_3\dot{X}_4-X_4\dot{X}_3=\frac{1}{2}\sin^2\gamma(\phi_+ +\phi_-)=q\gamma'\sin^2\gamma
\EE
where in the last steps of the above equations we have made use of (\ref{eqn_soln1}) and (\ref{eqn_soln2}) and hence $\dot{\gamma}=\frac{1}{2}(\gamma_+ +\gamma_-)
=\frac{1}{q}\phi'$ and $\dot{\phi}=\frac{1}{2}(\phi_+ +\phi_-)=q\gamma'$. This means, for finite nonzero $q$,
all of the above expressions are
total derivatives of functions of either $\gamma$ or $\phi$ w.r.t to $\sigma$ and thus $J_{AB}=0$.
 For $q=0$ or $q\to\infty$ one can also show that $J_{AB}=0$, using the fact that $\int_0^L d\sigma\gamma_+=\int_0^L
 d\sigma\Gamma'(\xi^+)=2\int_{\tau/2}^{\tau/2+L/2}d\xi^+\Gamma'(\xi^+)=0$ ((see the paragraph below
 equation (\ref{eqn_soln3}))), and similarly
 $\int_0^L d\sigma\gamma_-=0$, $\int_0^L d\sigma\phi_+=0$, and $\int_0^L d\sigma\phi_-=0$. In summary, the string solutions we consider do not carry any angular momentum.

\subsection{Higher charges of the solutions via B\"{a}cklund Transformation}
In this subsection we investigate the infinite number of local commuting charges generated via 
B\"{a}cklund transformation, following \cite{Ogielski:1979hv} and \cite{Arutyunov:2003rg}. To avoid confusion with the spherical coordinate,
we shall use $s$, instead of the commonly used $\gamma$, to denote the spectral parameter. 
The B\"{a}cklund transformation
is given by
\BE\label{eqn_BT}
& &2s^2\partial_+(X(s)+X)=(1+s^2)(X(s)\cdot\partial_+X)(X(s)-X)\nonumber\\
& &2\partial_-(X(s)-X)=-(1+s^2)(X(s)\cdot\partial_-X)(X(s)+X)
\EE
together with the normalization conditions
\BE\label{eqn_BT_normalization}
X(s=0)=X,\ \ X(s)\cdot X(s)=1,\ \ X(s)\cdot X=\frac{1-s^2}{1+s^2}
\EE
It is readily shown that $X(s)$ satisfies the same e.o.m.'s and constraints (\ref{eqn_o6_model}) as $X$.

Local conservation laws are derived by contracting the first equation of (\ref{eqn_BT}) with $\partial_-X$ and the
second equation with $\partial_+X$ and then adding the two equations, giving
\BE
\partial_\tau\left(s^2X(s)\cdot\partial_-X+X(s)\cdot\partial_+X\right)
+\partial_\sigma\left(s^2X(s)\cdot\partial_-X-X(s)\cdot\partial_+X\right)=0
\EE
which leads to the conserved quantity
\BE\label{eqn_Es}
{\cal E}(s)\equiv \int_0^L\frac{d\sigma}{2L}\left[sX(s)\cdot\partial_+X+s^3X(s)\cdot\partial_-X\right]
\EE
Expanding $X(s)$ as a power series in $s$
\BE
X(s)=\sum_{k=0}^\infty X^{(k)}s^k
\EE
we can use ${\cal E}(s)$ as a generating function of an infinite number of local charges ${\cal E}_k$
\BE
{\cal E}(s)=\sum_{k=2}^\infty{\cal E}_ks^k
\EE
where
\BE
{\cal E}_k=\int_0^L\frac{d\sigma}{2L}\left[X^{(k-1)}\cdot\partial_+X+X^{(k-3)}\cdot\partial_-X\right],\ \ k\ge 2
\EE
and $X^{(k)}=0$ for $k$ negative.

To evaluate the local charges it is preferable to have an exact solution for $X(s)$, when 
this is unavailable can find the solution as a power series expansion in $s$. Equations (3.16) to (3.18) of
\cite{Arutyunov:2003rg} give such an expression for a perturbative solution $X(s)=\sum X^{(k)}s^k$ using as their 
trial or bare solutions, $X$,  the folded or circular string. However, when the $X$'s are themselves continuous but have 
discontinuous
$\partial_+X$ and $\partial_-X$ (which
is the case for the kinky solutions that are of primary interest in this paper), that expression
is incorrect and gives discontinuous and even singular $X^{(k)}$ for higher $k$'s. We shall now illustrate this with
the $q=0$ case, which gives $\phi=0$, and where we will be able to find an exact solution 
to the B\"{a}cklund transformation.
Our solution corresponds to
\BE
X_1=\cos\gamma,\ \ X_3=\sin\gamma,\ \ X_2=X_4=X_5=X_6=0
\EE
and is simply the $O(2)$ sigma model. Now the normalization condition alone determines the solution. Let us make
the ansatz
\BE
X_1(s)=\cos\gamma(s),\ \ X_3(s)=\sin\gamma(s),\ \ X_2(s)=X_4(s)=X_5(s)=X_6(s)=0
\EE
then
\BE
X(s)\cdot X=\cos(\gamma(s)-\gamma)=\frac{1-s^2}{1+s^2}
\EE
Defining $s=\tan\frac{\beta}{2}$, we obtain an
\footnote{There is at least one other solution which can be obtained by letting $s\rightarrow -s$. 
It is clear that this is a solution as the B\"{a}cklund  equations are even in $s$. 
In our case it is straightforward to also find this solution and the charges are trivially related to 
those described below.} exact solution to the Backlund transformation
\BE
\gamma(s)=\gamma+\beta=\gamma+2\arctan(s)
\EE
giving
\BE
X_1(s)=\frac{1-s^2}{1+s^2}\cos\gamma-\frac{2s}{1+s^2}\sin\gamma,\ \ X_3(s)=\frac{1-s^2}{1+s^2}\sin\gamma
+\frac{2s}{1+s^2}\cos\gamma
\EE
It can be explicitly checked that the above $X(s)$ satisfies (\ref{eqn_BT}). It is clear that $X(s)$ corresponds
to a rotation of $X$ in the $13$ plane.
Hence we see that $X(s)$ indeed admits a power expansion around $s=0$: $X(s)=\sum_{k=0}^\infty X^{(k)}s^k$
with $X^{(k)}=\frac{1}{k!}\frac{d^kX(s)}{ds^k}|_{s=0}$, which gives
\BE
X_1^{(1)}=-2\sin\gamma,\ X_1^{(2)}=-2\cos\gamma, \ X_1^{(3)}=2\sin\gamma,\ ...
\EE
Let us compare this with the perturbative solution given by (3.16) to (3.18) of \cite{Arutyunov:2003rg}, which we denote as
$\tilde{X}^{(k)}$
\BE
& &\tilde{X}_1^{(1)}=\frac{2\partial_+X_1}{\kappa}=-2\sin\gamma\frac{\partial_+\gamma}{\kappa}\nonumber\\
& &\tilde{X}_1^{(2)}=\frac{2\partial_+^2X_1}{\kappa^2}=-2\cos\gamma\frac{(\partial_+\gamma)^2}{\kappa^2}-2\sin\gamma
\frac{\partial_+^2\gamma}{\kappa^2}=-2\cos\gamma-2\sin\gamma
\frac{\partial_+^2\gamma}{\kappa^2}
\EE
where in the second equation we have used the fact $(\partial_+\gamma)^2=\kappa^2$ (see (\ref{eqn_soln3})). We see
that $\tilde{X}_1^{(1)}$ differs from $X_1^{(1)}$ by the factor $\frac{\partial_+\gamma}{\kappa}$ (which alternates
between $+1$ and $-1$ over one period) and is thus discontinuous, $\tilde{X}_1^{(2)}$ differs from $X_1^{(2)}$ by a term involving
$\partial_+^2\gamma$ and thus exhibits delta-functional singularity. We omit the explicit expressions here,
but one can readily check that $\tilde{X}_1^{(3)}$ differs
from $X_1^{(3)}$ by the factor $\frac{\partial_+\gamma}{\kappa}$ as well as terms involving $\partial_+^2\gamma$ and
$\partial_+^3\gamma$ and is thus more singular. Similarly, higher and higher $\tilde{X}_1^{(k)}$'s contain
higher derivatives of $\gamma$ and exhibit worse singular behavior. Hence we see that $\tilde{X}^{(k)}_1$ cannot be
the right solution because the string should have a continuous profile order by order in $s$.
This illustrates a subtlety when solving the B\"{a}cklund transformation order by order in $s$: at each order there might be
more than one solution satisfying equation (\ref{eqn_BT}) (both $X^{(k)}_1$ and $\tilde{X}^{(k)}_1$ above satisfy the equation)
, and one has to choose the solution that gives a continuous string profile at each order (in our case it is $X^{(k)}_1$).

Let us now compute the generating function of local charges ${\cal E}(s)$ for the $q=0$ case. One readily finds, using the exact solution $X(s)$
we have
\BE
{\cal E}(s)=\int_0^L\frac{d\sigma}{2L}\left[sX(s)\cdot\partial_+X+s^3X(s)\cdot\partial_-X\right]
=\int_0^L\frac{d\sigma}{2L}\left(s\partial_+\gamma +s^3\partial_-\gamma\right)\sin\beta=0
\EE
where in the last step we have used the fact that $\int_0^L d\sigma \partial_+\gamma=\int_0^L d\sigma\Gamma'(\xi^+)
=2\int_{\tau/2}^{\tau/2+L/2}d\xi^+\Gamma'(\xi^+)=0$
(see the paragraph under (\ref{eqn_soln3})) and similarly $\int_0^L d\sigma \partial_-\gamma=0$.
Hence we see that all the higher charges ${\cal E}_k$ vanish. The $q=\infty$ case is completely analogous.

The above construction can be generalized to the case of nonzero and finite $q$. We make the ansatz
\BE
X_1(s)=\cos\gamma(s)\cos \phi(s),& & \ X_2(s)=\cos\gamma(s)\sin\phi(s),\ \ \ X_3(s)=\sin\gamma(s)\cos\phi(s),\nn\\
   & &\kern-30pt  X_4(s)=\sin\gamma(s)\sin \phi(s),\ \ X_5(s)=X_6(s)=0
\EE
and from the normalization condition we have,
\BE
X(s)\cdot X &=& \cos(\gamma(s)-\gamma)\cos (\phi(s)-\phi)\nn\\
            &=& \frac{1-s^2}{1+s^2}.
\EE
We further assume that $\gamma(s)=\gamma+a(s)$ and $\phi(s)=\phi+b(s)$, where $a,b$ are independent of $(\tau,\sigma)$.
Then the normalization condition becomes
\BE
\cos a\cos b=\frac{1-s^2}{1+s^2}
\EE
Also notice
\BE\label{eqn_Xs_dot_partial_X}
& &X(s)\cdot\partial_+ X=\gamma_+\sin a\cos b+\phi_+\cos a\sin b\nonumber\\
& &X(s)\cdot\partial_- X=\gamma_-\sin a\cos b+\phi_-\cos a\sin b
\EE
Using the facts $s^2=\frac{1-\cos a\cos b}{1+\cos a\cos b}$, $\phi_+=q\gamma_+$, $\phi_-=-q\gamma_-$, and
going through a series of trigonometry identities, we can simplify the B\"{a}cklund transformation (\ref{eqn_BT})
and find that it is
satisfied when
\BE\
q\sin a=\sin b
\EE
For example, the first equation in (\ref{eqn_BT}) for $A=1$ becomes
\BE
[\sin \gamma\sin\phi(s)-\sin\gamma(s)\sin\phi](q\sin a-\sin b)=0
\EE
Then using equation (\ref{eqn_Xs_dot_partial_X}) and the fact $\int_0^Ld\sigma\gamma_\pm=0, \int_0^Ld\sigma\phi_\pm=0$,
we conclude that ${\cal E}(s)$ defined in (\ref{eqn_Es}) vanishes identically.

In summary, we have found the exact solution to the B\"{a}cklund transformation, which corresponds to shifting $\gamma$ and $\phi$ by
$a(s)$ and $b(s)$ respectively, with $a,b$ determined by $\cos a\cos b=\frac{1-s^2}{1+s^2}$ and $q\sin a=\sin b $. The
generating function ${\cal E}(s)$ turns out to always vanish, hence so do all the local charges ${\cal E}_k$.

\subsection{Higher Charges via Monodromy Matrix}\label{subsection_monodromy}
A second method for constructing conserved charges of an integrable systems is to expand the eigenvalues of the monodromy matrix in powers of the spectral parameter about its singular points. We shall work in the formulation of $SU(2)$ principal chiral model following \cite{Kazakov:2004qf}. The $SU(2)$ principal chiral field is
\BE
g=
\begin{pmatrix}
X_1+iX_2& X_3+iX_4\\
-X_3+iX_4& X_1-iX_2
\end{pmatrix}
=
\begin{pmatrix}
Z_1&Z_2\\
-\bar{Z}_2&\bar{Z}_1
\end{pmatrix}\in SU(2)
\EE
The $su(2)$-algebra valued right currents are defined as $j=-g^{-1}dg$ (notice that our definition of $j$ differs from that of
\cite{Kazakov:2004qf} by an overall sign), which can be shown to satisfy
\BE\label{eqn_j_pm}
\partial_+j_-+\partial_-j_+=0,\ \ \partial_+j_--\partial_-j_+ -[j_+,j_-]=0
\EE
One can define a one-parameter family of currents $J(x)$
\BE
J_\pm(x)=\frac{j_\pm}{1\mp x}
\EE
which has vanishing curvature
\BE
\partial_+J_- - \partial_-J_+ -[J_+,J_-]=0
\EE
due to eqn.'s (\ref{eqn_j_pm}). Let us consider the Wilson line defined as
\BE
\Omega(\sigma_1,\sigma_2)\equiv P e^{\int_{\sigma_2}^{\sigma_1}d\sigma J_{\sigma}(\sigma)}
=P e^{\int_{\sigma_2}^{\sigma_1}d\sigma \frac{1}{2}(\frac{j_+}{1-x}-\frac{j_-}{1+x})}
\EE
where $\sigma_1>\sigma_2$ and $P$ stands for path ordering, putting larger values of $\sigma$ on the left. $\Omega(\sigma_1,
\sigma_2)$ as defined above has the properties
\BE
\partial_{\sigma_1}\Omega(\sigma_1,\sigma_2)=J_\sigma(\sigma_1)\Omega(\sigma_1,\sigma_2)
\EE
and
\BE
\Omega(\sigma_1,\sigma_2)\Omega(\sigma_2,\sigma_3)=\Omega(\sigma_1,\sigma_3),\ \ \sigma_1>\sigma_2>\sigma_3
\EE
The monodromy matrix is the Wilson loop $\Omega(L,0)$. The vanishing of the curvature of $J$ implies that under
time evolution the monodromy matrix only undergoes conjugation, hence its eigenvalues are conserved quantities.

For the solutions constructed in Section \ref{section_our_solution},
\BE
g=\begin{pmatrix}
\cos\gamma e^{i\phi}&\sin\gamma e^{i\phi}\\
-\sin\gamma e^{-i\phi}&\cos\gamma e^{-i\phi}
\end{pmatrix}
\EE
and
\BE
j=-i\left[\sigma^1(\sin 2\gamma)d\phi+\sigma^2d\gamma+\sigma^3(\cos 2\gamma)d\phi\right]
\EE
with $\sigma^{1,2,3}$ being the standard Pauli matrices. Let us first consider the simplest case of $q=0$ and
$q=\infty$. In the $q=0$ case, $\gamma\neq 0$ and $\phi=0$. This leads to $j=-i\sigma^2d\gamma$, and $J_\sigma
=\frac{1}{2}\left(\frac{\gamma_+}{1-x}-\frac{\gamma_-}{1+x}\right)(-i\sigma^2)$. Since $J_\sigma$'s at different values
of $\sigma$ commute, no path ordering is necessary. This gives
\BE
\Omega(\sigma_1,\sigma_2)=e^{\int_{\sigma_2}^{\sigma_1}d\sigma
 \frac{1}{2}\left(\frac{\gamma_+}{1-x}-\frac{\gamma_-}{1+x}\right)(-i\sigma^2) }
\EE
Due to the fact $\int_0^L d\sigma\gamma_\pm=0$, the monodromy matrix $\Omega(L,0)={\bf 1}_{2\times 2}$ in this case. The
$q=\infty$ case is similar. Now $j=-i\sigma^3d\phi$, $J_\sigma=\frac{1}{2}\left(\frac{\phi_+}{1-x}-\frac{\phi_-}{1+x}
\right)(-i\sigma^3)$, and
\BE
\Omega(\sigma_1,\sigma_2)=e^{\int_{\sigma_2}^{\sigma_1}d\sigma
 \frac{1}{2}(\frac{\phi_+}{1-x}-\frac{\phi_-}{1+x})(-i\sigma^3) }
\EE
which again leads to $\Omega(L,0)={\bf 1}_{2\times 2}$, using the fact $\int_0^L d\sigma\phi_\pm=0$.

Next let us consider a general nonzero finite value of $q$. Using the facts $\phi_+=q\gamma_+,\phi_-=-q\gamma_-$, we find
\BE
j_\pm=\mp i\gamma_\pm\left(\sigma^1 q\sin 2\gamma\pm\sigma^2+\sigma^3q\cos 2\gamma\right)
\EE
and hence
\BE\label{eqn_J_sigma}
J_\sigma=-i\left(\sigma^1qf\sin2\gamma+\sigma^2w+\sigma^3qf\cos2\gamma\right)
\EE
where we have defined the functions $f,w$ as
\BE
f=\frac{1}{2}\left(\frac{\gamma_+}{1-x}+\frac{\gamma_-}{1+x}\right),\ \
w=\frac{1}{2}\left(\frac{\gamma_+}{1-x}-\frac{\gamma_-}{1+x}\right)
\EE
One salient feature of our solutions is the discontinuities exhibited in $\gamma_\pm$(see the paragraph below
 equation (\ref{eqn_soln3})). Thus we divide the string into segments. In each segment, $\gamma_\pm$ are constant, and so
 $\partial_\sigma\gamma=\frac{1}{2}(\gamma_+
-\gamma_-)$ is also constant. This means that in each segment $f,w$ are
constant and the $\sigma$-dependence of $J_\sigma$ comes purely from $\gamma$, which is a linear function of $\sigma$.
Going from one segment to the next, $\gamma_+$, $\gamma_-$, or both, jump and consequently $f,w$ and $\partial_\sigma\gamma$ jump from one
value to another. This suggests we should compute the monodromy matrix in a piecewise manner
\BE
\Omega(L,0)=\Omega(\sigma_0=L,\sigma_1)\Omega(\sigma_1,\sigma_2)...\Omega(\sigma_{m-1},\sigma_m=0)
\EE
where $\sigma_1,\sigma_2,...,\sigma_{m-1}$ are values of $\sigma$ at which $\gamma_+$, $\gamma_-$, or both, jump. Let us
solve
\BE
\partial_\sigma\Omega(\sigma,\sigma_{k+1})=J_\sigma(\sigma)\Omega(\sigma,\sigma_{k+1}),\ \ \sigma_{k+1}<\sigma<\sigma_k
\EE
for $k=0,1,...,m-1$. This can be done by defining the auxiliary 2-component ``wave-function'' $\Psi(\sigma)$
\BE
\Psi(\sigma)\equiv\Omega(\sigma,\sigma_{k+1})\Psi(\sigma_{k+1})
\EE
with $\Psi(\sigma_{k+1})$ being an arbitrary constant 2-component vector. We see that
\BE
\partial_\sigma\Psi(\sigma)=J_\sigma(\sigma)\Psi(\sigma)
\EE
Given the expression of $J_\sigma$ in (\ref{eqn_J_sigma}) and the fact that $f,w,\partial_\sigma\gamma$ are all constant
in the interval $(\sigma_{k+1},\sigma_k)$, this problem is simply a textbook one of solving the Schrodinger equation for
a spin-$1/2$ particle in a rotating magnetic field. Using the identity $\sigma^3\cos 2\gamma+\sigma^1\sin 2\gamma
=e^{-i\gamma\sigma^2}\sigma^3 e^{i\gamma\sigma^2}$ and going to the rotating frame $\Psi=e^{-i\gamma\sigma^2}\tilde{\Psi}$,
one readily solves this Schrodinger equation and finds
\BE
\Omega(\sigma,\sigma_{k+1})=e^{-i\gamma(\sigma)\sigma^2}e^{-i\left\{(w^{(k)}-\partial_\sigma\gamma^{(k)})\sigma^2
+qf^{(k)}\sigma^3 \right\}(\sigma-\sigma_{k+1})}e^{i\gamma(\sigma_{k+1})\sigma^2}
\EE
where we have added the superscript $(k)$ to $f,w,\partial_\sigma\gamma$ to make it explicit that these are their (constant)
values in the segment $\sigma_{k+1}<\sigma<\sigma_k$. Thus we have
\BE
\Omega(\sigma_k,\sigma_{k+1})=e^{-i\gamma(\sigma_k)\sigma^2}e^{-i\left\{(w^{(k)}-\partial_\sigma\gamma^{(k)})\sigma^2
+qf^{(k)}\sigma^3 \right\}(\sigma_k-\sigma_{k+1})}e^{i\gamma(\sigma_{k+1})\sigma^2}
\EE
and in turn
\BE
& &\Omega(L,0)=\nonumber\\
& &e^{-i\gamma(\sigma_0)\sigma^2}e^{-i\left\{(w^{(0)}-\partial_\sigma\gamma^{(0)})\sigma^2
+qf^{(0)}\sigma^3 \right\}(\sigma_0-\sigma_1)}...e^{-i\left\{(w^{(m-1)}-\partial_\sigma\gamma^{(m-1)})\sigma^2
+qf^{(m-1)}\sigma^3 \right\}(\sigma_{m-1}-\sigma_m)}e^{i\gamma(\sigma_m)\sigma^2}\nonumber\\
\EE
One final simplification comes from the observation that
\BE
w-\partial_\sigma\gamma=\frac{1}{2}\left(\frac{\gamma_+}{1-x}-\frac{\gamma_-}{1+x}\right)-\frac{1}{2}(\gamma_+-\gamma_-)
=x\frac{1}{2}\left(\frac{\gamma_+}{1-x}+\frac{\gamma_-}{1+x}\right)=xf
\EE
which implies
\BE
(w^{(k)}-\partial_\sigma\gamma^{(k)})\sigma^2
+qf^{(k)}\sigma^3=f^{(k)}(x\sigma^2+q\sigma^3),
\EE
which commute between different segments. Using this fact, we find
\BE
\Omega(L,0)=e^{-i\gamma(\sigma_0)\sigma^2}e^{-i(x\sigma^2+q\sigma^3)\sum_{k=0}^{m-1}f^{(k)}(\sigma_k-\sigma_{k+1})}e^{i\gamma(\sigma_m)\sigma^2}
\EE
and we note that
\BE
\sum_{k=0}^{m-1}f^{(k)}(\sigma_k-\sigma_{k+1})=\int_0^L d\sigma f=\int_0^L d\sigma \frac{1}{2}\left(\frac{\gamma_+}{1-x}+\frac{\gamma_-}{1+x}\right)
=0
\EE
using $\int_0^L d\sigma\gamma_\pm=0$. Also noticing that $\gamma(\sigma_0=L)=\gamma(\sigma_m=0)$, we finally conclude that
\BE
\Omega(L,0)={\bf 1}_{2\times 2}
\EE
In summary, we have shown that the monodromy matrix is simply the identity matrix for all the solutions constructed
in Section \ref{section_our_solution}.

Now that we have shown for the kinky solutions the monodromy matrix is trivial, and hence the quasi-momentum $p(x)$ defined by
$\text{tr}\Omega(L,0;x)=2\cos p(x)$ is identically zero (modulo $2\pi m$, $m$ being an integer), we would like
to comment on the previous constructions which found that the quasi-momentum has 
poles at $x=\pm 1$ with the residue being $-\pi\kappa$. The argument
behind this claim is as follows (see e.g. \cite{Beisert:2004ag}). One can diagonalize $j_\pm$ to $j_\pm^{diag}$, then
as $x\to\pm 1$ we can drop the path ordering and
\BE
& &\text{tr}\Omega(L,0;x)\to\text{tr}\left(e^{\int_0^L d\sigma\frac{\pm 1}{2}\frac{j_\pm^{diag}}{1\mp x}}\right)=
\text{tr}\begin{pmatrix}exp\left(\frac{\pm 1}{2(1\mp x)}\int_0^L d\sigma\ i\lambda\right)&0\\
0&exp\left(\frac{\pm 1}{2(1\mp x)}\int_0^L d\sigma\ (-i\lambda)\right)\end{pmatrix}\nonumber\\
& &=2\cos\left(\frac{1}{2(1\mp x)}\int_0^L d\sigma\ \lambda\right)
\EE where we have written \BE\label{eqn_diag_j_pm}
j_\pm^{diag}=\begin{pmatrix}i\lambda&0\\ 0&-i\lambda\end{pmatrix}
\EE This implies that $p(x)$ has a pole at $x=\pm 1$ with the
residue being $-\frac{1}{2}\int_0^L d\sigma\ \lambda$. Meanwhile
notice that \BE\label{eqn_j_pm_squared}
\text{tr}(j_\pm^2)=-2\partial_+X_A\partial_+X_A=-2\kappa^2 \EE by
virtue of the conformal gauge constraint, which implies
$\lambda^2=\kappa^2$. The  argument is then that $\lambda$
is a $\sigma$-independent constant, either $\kappa$ or $-\kappa$,
over the entire string length and this then gives a residue
$-\frac{L\kappa}{2}=-\pi\kappa$ (setting $L=2\pi$). However, this
construction excludes the case when $j=-g^{-1}dg$ is not
continuous (in fact it does not have to be continuous, since it involves first derivatives of string
embedding $X(\tau,\sigma)$). We can have step functions in $j$ and
still get a continuous string profile $X$, which seems to be a sensible solution. 
This means $\lambda$ does not have to be a
$\sigma$-independent constant: it can jump between $\kappa$ and
$-\kappa$ as we move from one segment of the string to another. This
is exactly what is happening for the kinky string. Take the
$q=\infty$ case for example (in this case,
$j_\pm=-i\sigma^3\phi_\pm$ is already diagonal). We see that the
eigenvalue $\lambda=-\phi_\pm(\sigma)$, which can have an arbitrary
pattern of jumps and integrates to zero over the whole string
length, giving a vanishing residue. The moral of the story for the
$q=0$ and general nonzero finite $q$ cases is the same.

Given that the monodromy matrix for our solutions is always trivial,
it would seem that the usual local and non-local charges which are
found by expanding in powers of the spectral parameter are all vanishing. 
It is interesting to see if we can find other, non-trivial, charges.  The class of 
local charges constructed for principal chiral models based on the invariant tensors 
of the underlying algebra (see for example
  \cite{Goldschmidt:1980wq}
,\cite{Evans:1997xu},\cite{Evans:2000hx}) are one possibility. In particular, it is easy 
to see from the equations of motion (\ref{eqn_j_pm}) that
\BE
\partial_-Tr\left(j_+^m\right)=\partial_+Tr\left(j_-^m\right)=0
\EE for all $m$. These classically conserved charges are in
involution with the non-local Yangian symmetries of the principal
chiral model and indeed may survive to the quantum theory.
Unfortunately for us while they are not all zero they are
essentially trivial.This can be seen as follows: taking the
diagonalized form of $j_\pm$ as in (\ref{eqn_diag_j_pm}), we see
that for odd $m$ the traces $Tr(j_\pm^m)$ vanish, while for even $m$
we get $Tr(j_\pm^m)=2(-1)^{m/2}\kappa^m$ (using the fact
$\lambda^2=\kappa^2$, see (\ref{eqn_j_pm_squared})), which are the
same for all of our solutions. However, as mentioned in
\cite{Evans:1997xu} it is possible to consider differentials of these
charges, for example it is clear from equation (\ref{eqn_j_pm}) that
\BE Tr\left(j_+^m\right)\partial_+^pTr\left(j_+^q\right) \ {\rm{and}}\ 
Tr\left(j_-^m\right)\partial_-^pTr\left(j_-^q\right) \EE are conserved. Of course as the traces
are vanishing or constant we might suppose that these charges are also vanishing. However 
given the kinky feature of our solutions there is an ambiguity involved 
in their definition. Let us evaluate a simple case, 
$Tr\left(j_+\partial_+j_+\right)$, on
our solution, 
\BE
j_+&=&-iq \gamma_+\begin{pmatrix}\cos 2 \gamma&\sin 2\gamma-\frac{i}{q}\\
                                 \sin 2\gamma+\frac{i}{q}&-\cos 2\gamma\end{pmatrix}.
\EE
For our solutions  $\partial_+^2 \gamma=0$ everywhere except at the kinks where
\BE
\partial_+^2\gamma=\frac{\pm 2 \kappa}{\sqrt{1+q^2}}\delta\left(\xi^+_{\rm{kink}}\right)
\EE
with the sign corresponding to whether the value of $\gamma_+$ increases or decreases as we 
cross the kink.
We thus find the conserved charge 
\BE
Q&=&\int d\sigma\ Tr\left(j_+\partial_+j_+\right)\nn\\
&=&-4 \kappa \sqrt{1+q^2}\sum_{\rm{kinks}}\pm \gamma_+|_{\rm{kinks}}.
\EE
This leads us to the question of what exactly is the value of $\gamma_+$ at the kinks? In order for the
Virasoro constraints to be satisfied it must be $\frac{\pm \kappa}{\sqrt{1+q^2}}$ 
however the sign is somewhat arbitrary. If we choose it to be always positive we find 
the above charge to be vanishing for an even number of kinks. However if we choose the sign of 
$\gamma_+$ at each kink to correspond to that of $\partial_+^2 \gamma$ we get a non-vanishing result, which, while it is independent of $q$ does depend on the number of kinks. Currents involving more 
derivatives will result in more singular
charges while currents with more powers of $j_\pm$ result in simple powers of the overall coefficient.

\subsection{Example Solutions}\label{subsection_example_solutions}
In this section we give some example solutions which have $n$ evenly spaced jumps in the worldsheet first 
derivatives.

Example {\bf I}

\begin{figure}[ht]
\epsfig{figure=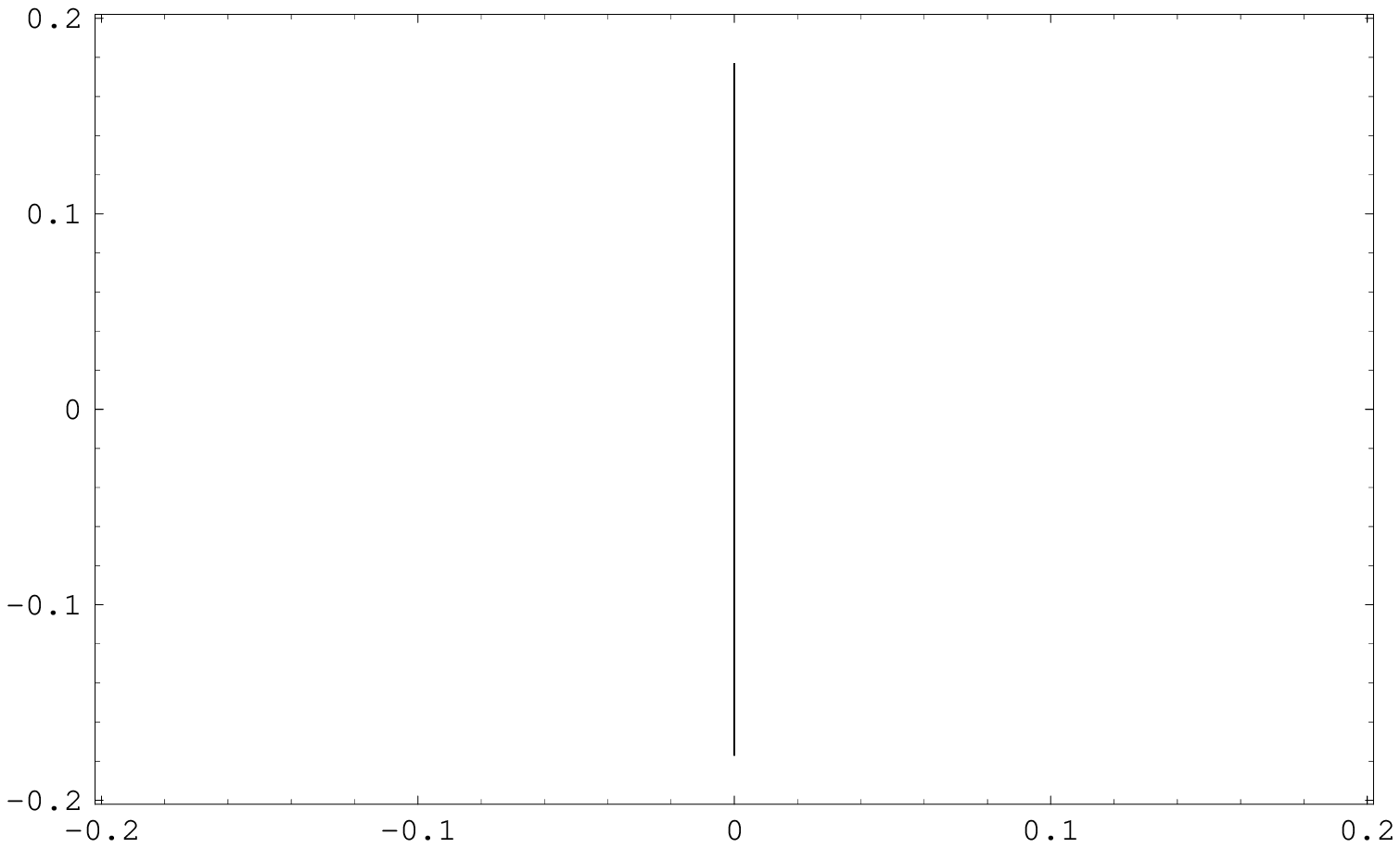,width=3.0in,angle=0}
\epsfig{figure=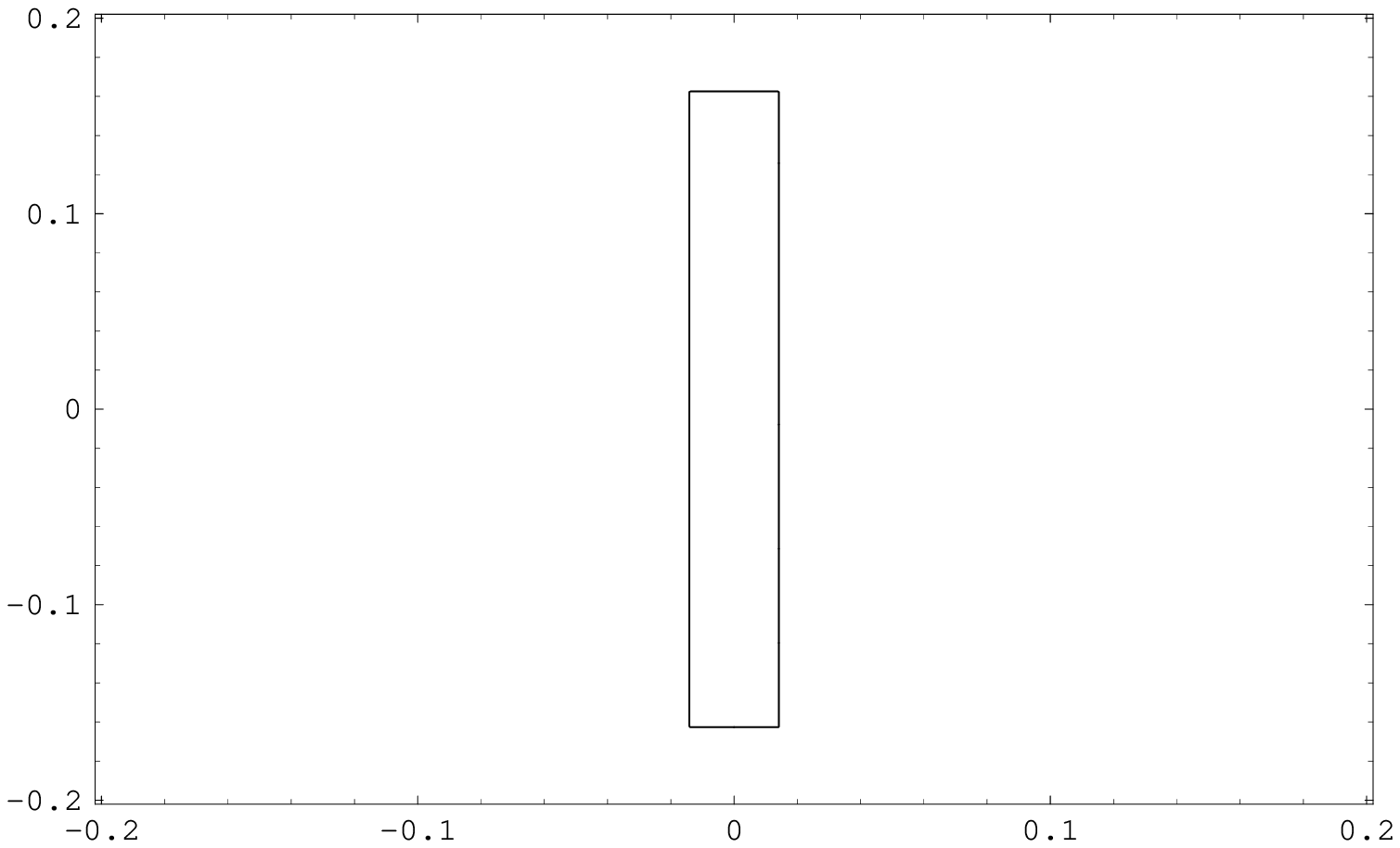,width=3.0in,angle=0}
\epsfig{figure=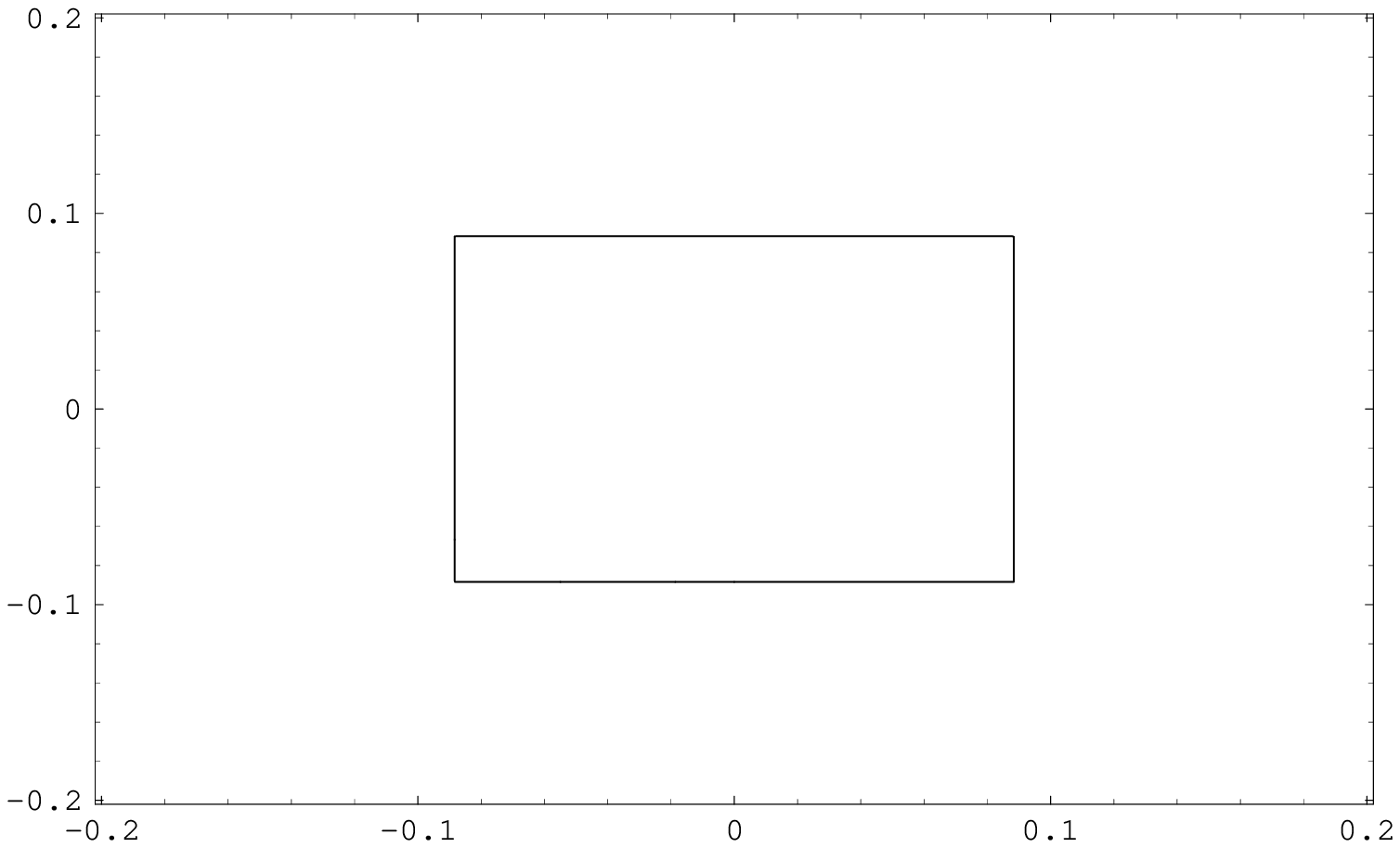,width=3.0in,angle=0}
\epsfig{figure=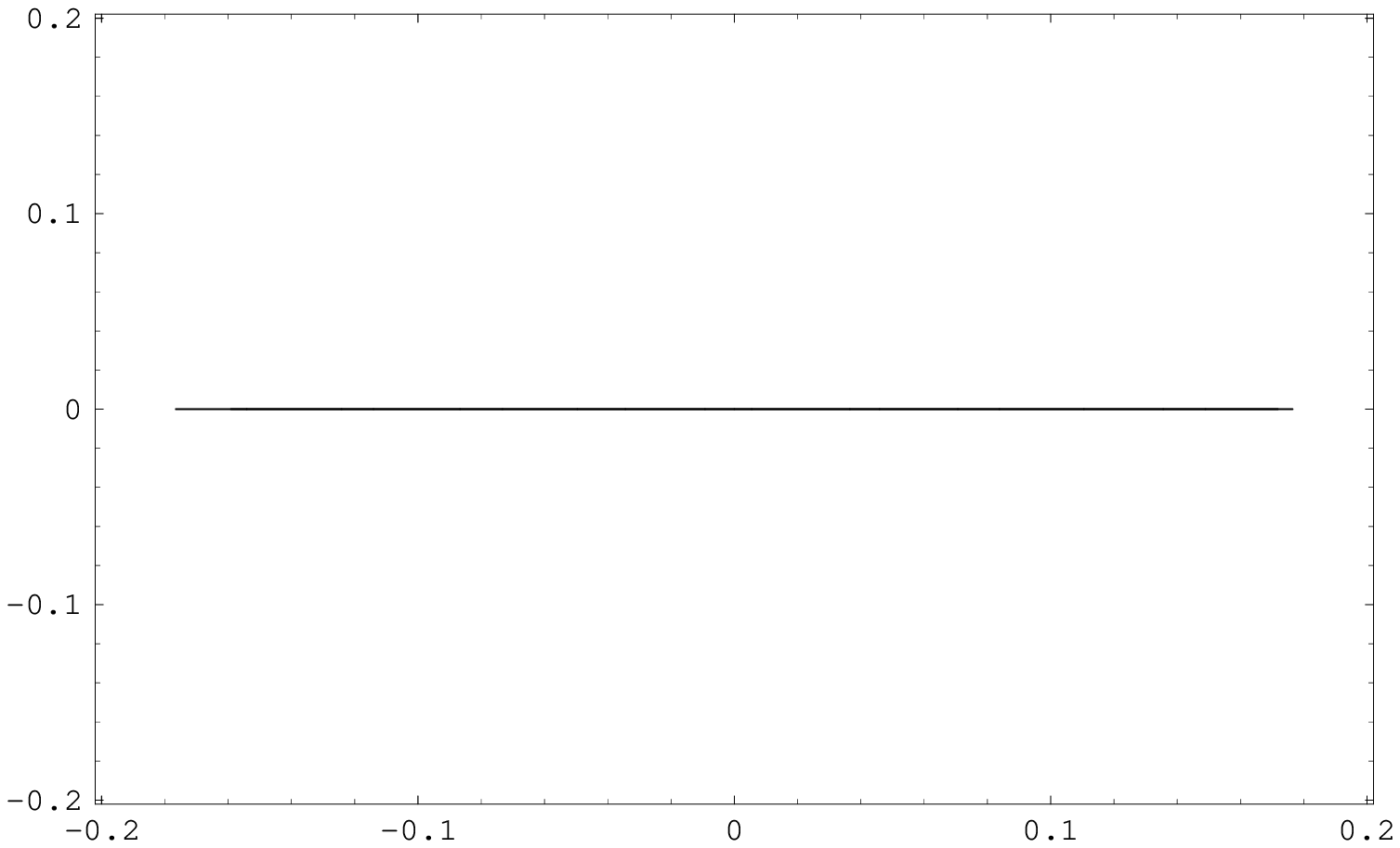,width=3.0in,angle=0}
\epsfig{figure=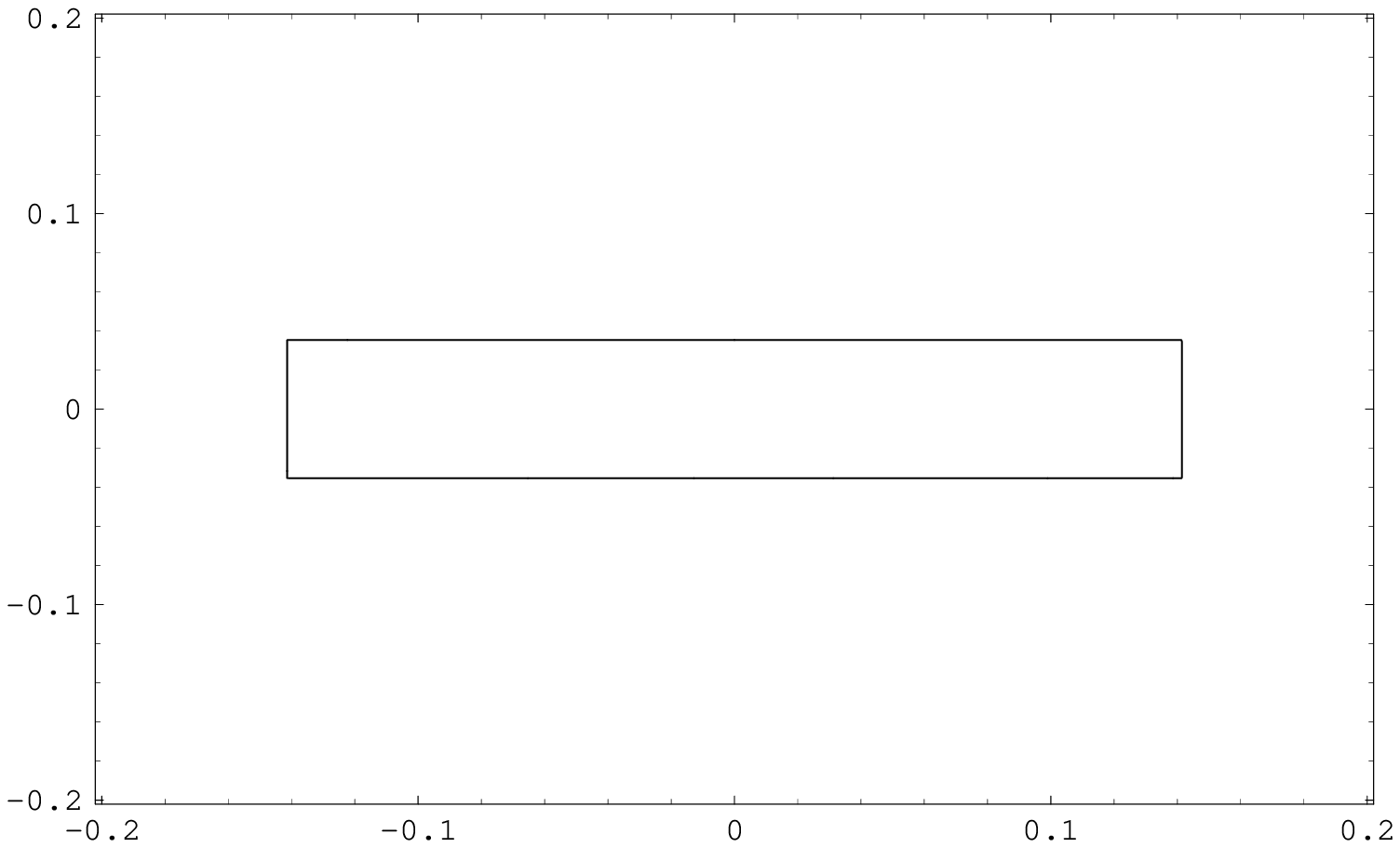,width=3.0in,angle=0}
\epsfig{figure=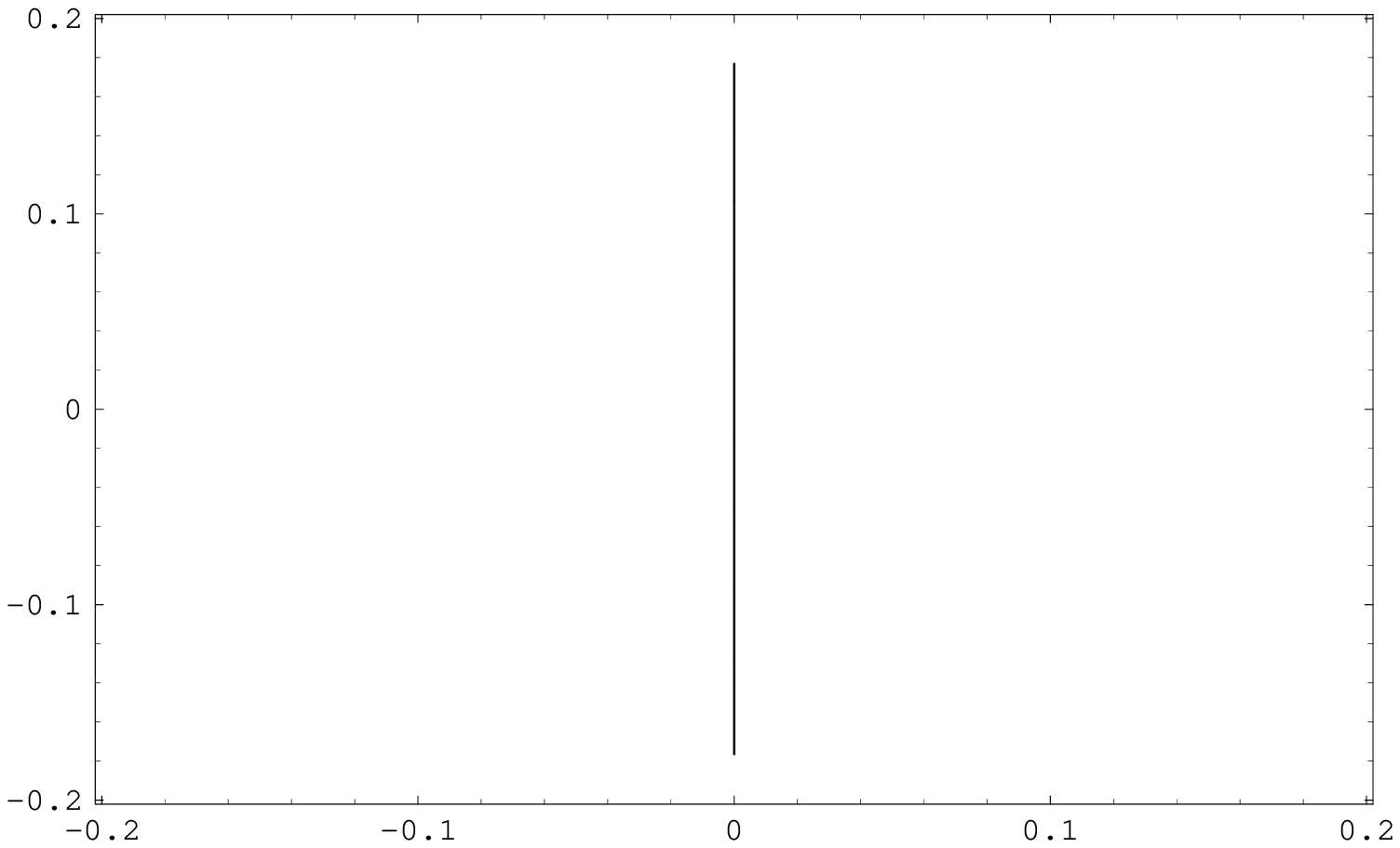,width=3.0in,angle=0}
\caption{The pulsating rectangle solution for $q=1$, at $\tau=0,0.02L$(first row, left to right),
$\tau=0.125,0.25$(second row), and $\tau=0.3,0.5$(third row);
horizontal axis is $\phi$ and vertical axis is $\gamma$, both labeled in units of $\kappa L$}
\label{figure_n1}
\end{figure}

Take $\Gamma'(\xi^+)=f(\xi^+)$ and $\tilde{\Gamma}'(\xi^-)=f(\xi^-)$ with
\BE
\label{feqn}
& &f(\xi)=\frac{\kappa}{\sqrt{1+q^2}} \ \ \ \ \ \ \text{  for } \frac{mL}{2}<\xi<\frac{mL}{2}+\frac{L}{4}\nonumber\\
& &f(\xi)=-\frac{\kappa}{\sqrt{1+q^2}} \ \ \ \ \ \ \text{  for } \frac{mL}{2}+\frac{L}{4}<\xi<\frac{mL}{2}+\frac{L}{2}
\EE ($m\in{\bf Z}$)
This leads to
\BE
\gamma=F(\xi^+)+F(\xi^-),\ \ \ \phi=q\left[F(\xi^+)-F(\xi^-)\right]
\EE
with $F(\xi)=\int_{L/8}^\xi f(\tilde{\xi})d\tilde{\xi}$. The time evolution of the string's profile is as follows:
at $\tau=0$, the string is folded and lies on the $\gamma$-axis, extending from
$\gamma=-\frac{\kappa}{\sqrt{1+q^2}}\frac{L}{4}$ to $\gamma=\frac{\kappa}{\sqrt{1+q^2}}\frac{L}{4}$;
at $\tau=\frac{L}{8}$, the string is a rectangle, with its four vertices being
$(\gamma,\phi)=(\pm\frac{\kappa}{\sqrt{1+q^2}}\frac{L}{8}, \pm q\frac{\kappa}{\sqrt{1+q^2}}\frac{L}{8})$;
at $\tau=\frac{L}{4}$, the string is folded and lies on the $\phi$-axis, extending from
$\phi=-q \frac{\kappa}{\sqrt{1+q^2}}\frac{L}{4}$ to $\phi=q \frac{\kappa}{\sqrt{1+q^2}}\frac{L}{4}$.
See Figure \ref{figure_n1} (Plots made using Mathematica$^\circledR$).
When we take the special case of $q=0$, we get a folded string pulsating on the $\gamma$-axis; when we take $q\to\infty$,
we get a folded string pulsating on the $\phi$-axis.

We can modify the solution by introducing a constant ``phase shift'' $\Delta$, i.e. taking $\Gamma'(\xi^+)=f(\xi^+)$
and $\tilde{\Gamma}'(\xi^-)=f(\xi^-+\Delta)$ with $f(\xi)$ the same as before. As can be seen, the solution obtained
this way is related to the old solution via
\BE
\gamma_{new}(\tau,\sigma)=\gamma_{old}(\tau+\Delta,\sigma-\Delta),
\ \ \phi_{new}(\tau,\sigma)=\phi_{old}(\tau+\Delta,\sigma-\Delta)
\EE
which means that adding the phase shift merely corresponds to a shift in $\tau$.
\\
\\
Example {\bf II}

\begin{figure}[ht]
\epsfig{figure=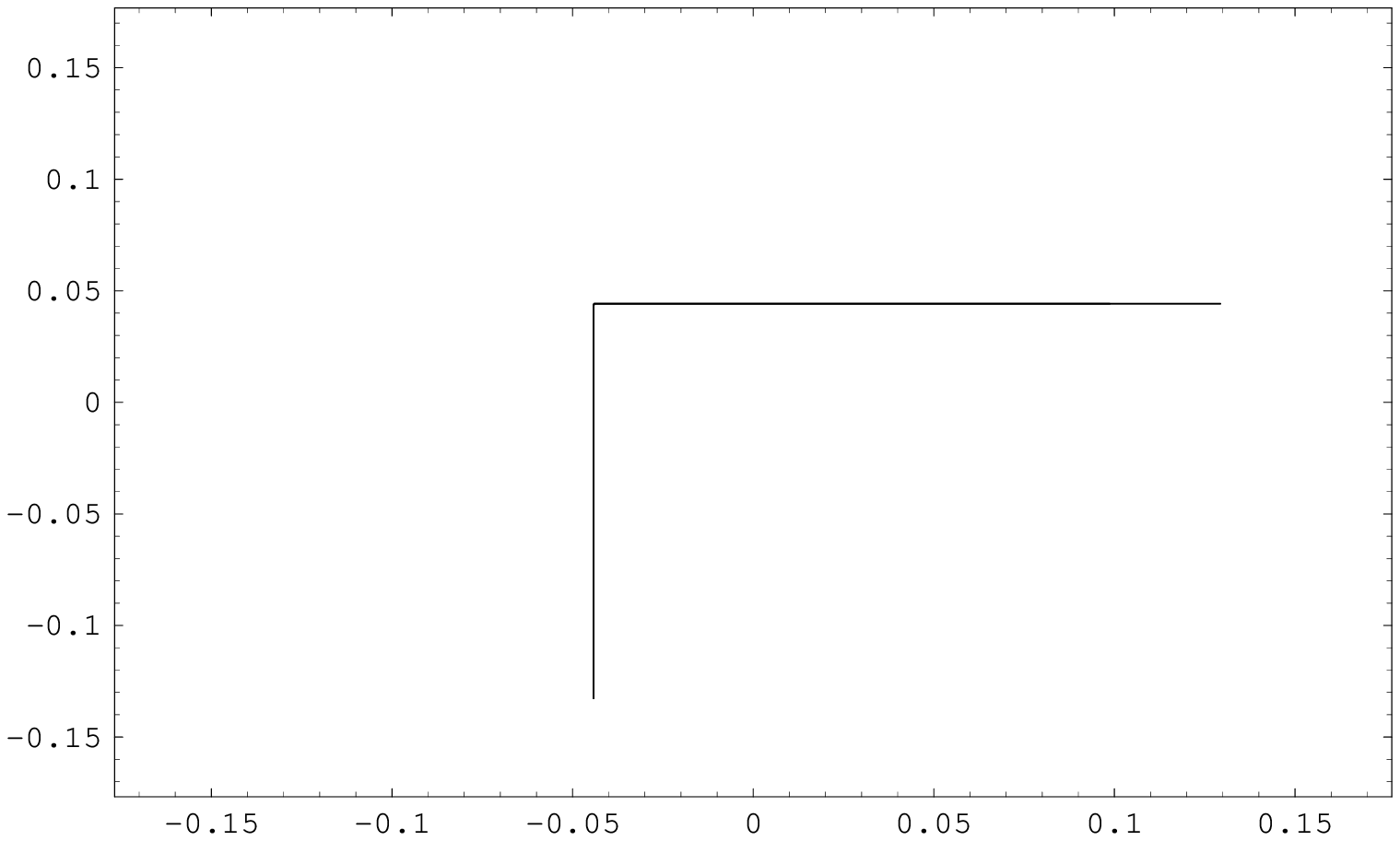,width=3.0in,angle=0}
\epsfig{figure=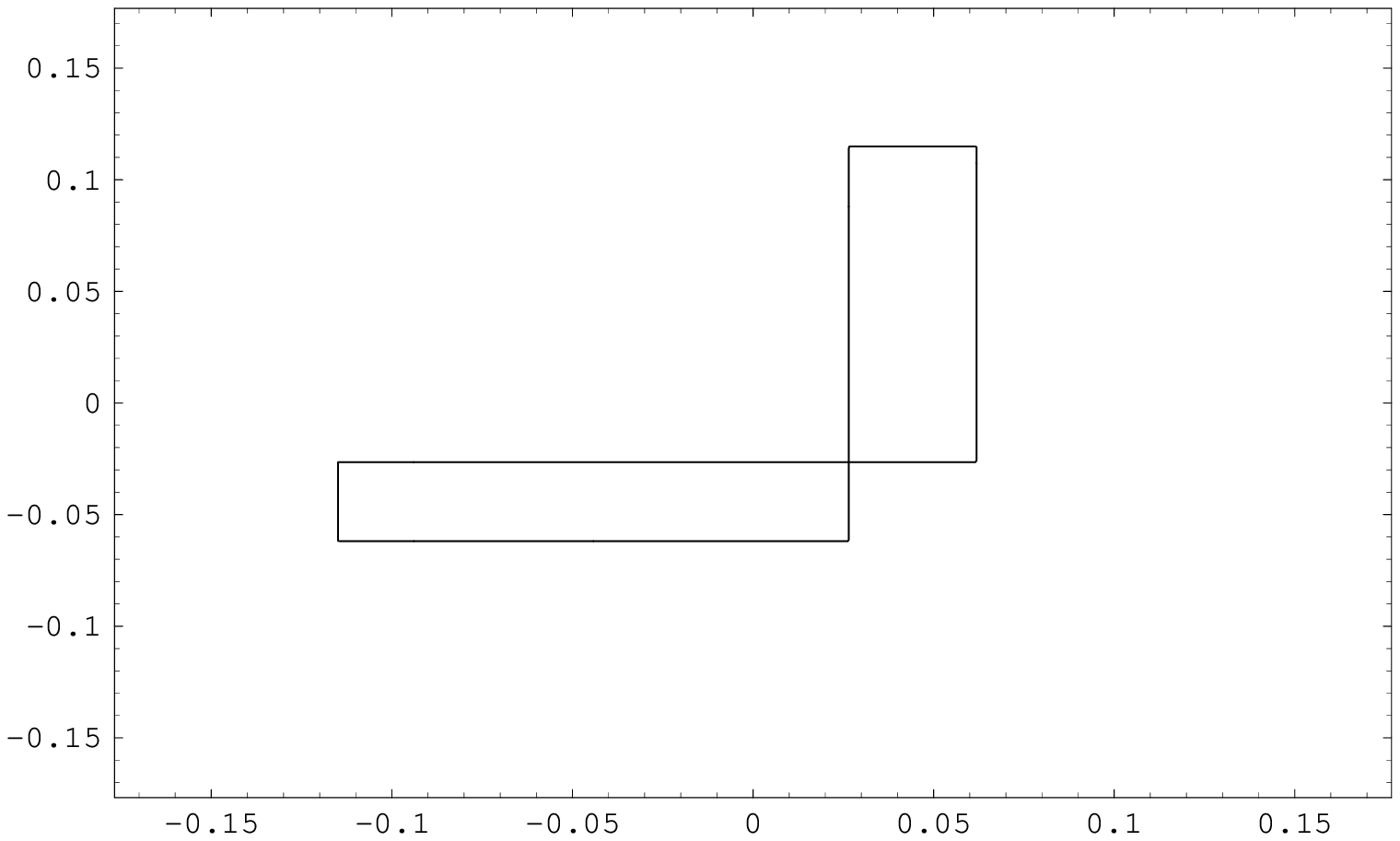,width=3.0in,angle=0}
\caption{The string solution for $n=2$, $q=1$, at $\tau=0$(left figure) and $\tau=0.1L$(right figure);
the axes(horizontal axis: $\phi$; vertical axis: $\gamma$) are labeled in units of $\kappa L$}
\label{figure_selfcrossing_n2}
\end{figure}

Let us generalize the above example by introducing an additional parameter $n$. Take
\BE
\Gamma'(\xi^+)=f(\xi^+),\ \ \tilde{\Gamma}'(\xi^-)=f(n\xi^-)
\EE
with $n$ being a positive integer, and $f(\xi)$ the same as before. This leads to
\BE
\gamma=F(\xi^+)+\frac{1}{n}F(n\xi^-),\ \ \phi=q\left[F(\xi^+)-\frac{1}{n}F(n\xi^-)\right]
\EE
Taking the special case of $n=1$ we get back example {\bf I}. For $n>1$, in general we get self-crossing strings. See
Figure \ref{figure_selfcrossing_n2} for the $n=2$ case.
One thing worth pointing out is, when $n>1$ the string never shrinks to a point, not even in the special cases of $q=0,\infty$ (while when $n=1$ the string shrinks to a point at
$\tau=\frac{L}{4}$ for $q=0$, and at $\tau=0$ for $q=\infty$), due to the $\frac{1}{n}$ factor in front of the right moving part $F(n\xi^-)$.

\begin{figure}[ht]
\epsfig{figure=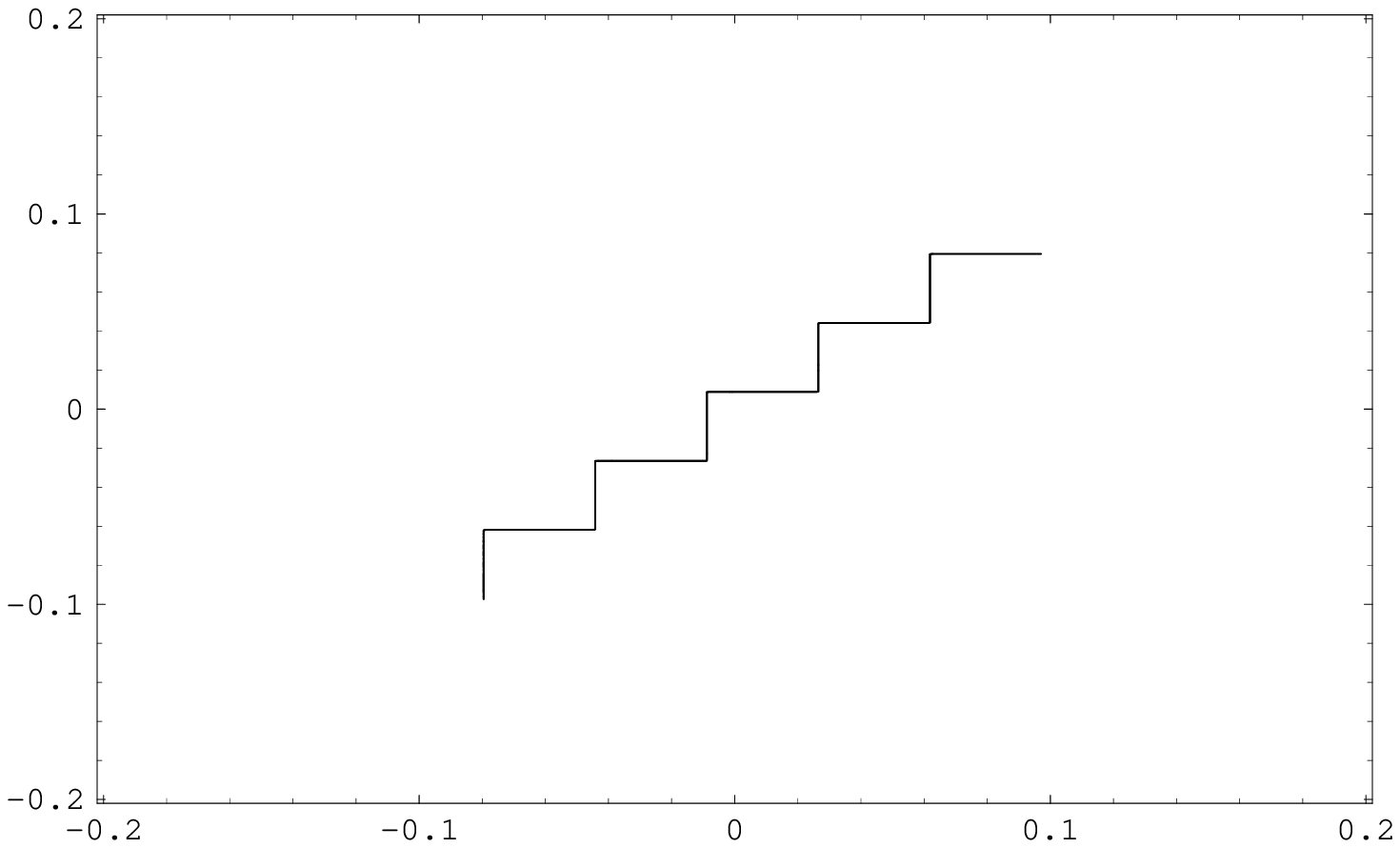,width=3.0in,angle=0}
\epsfig{figure=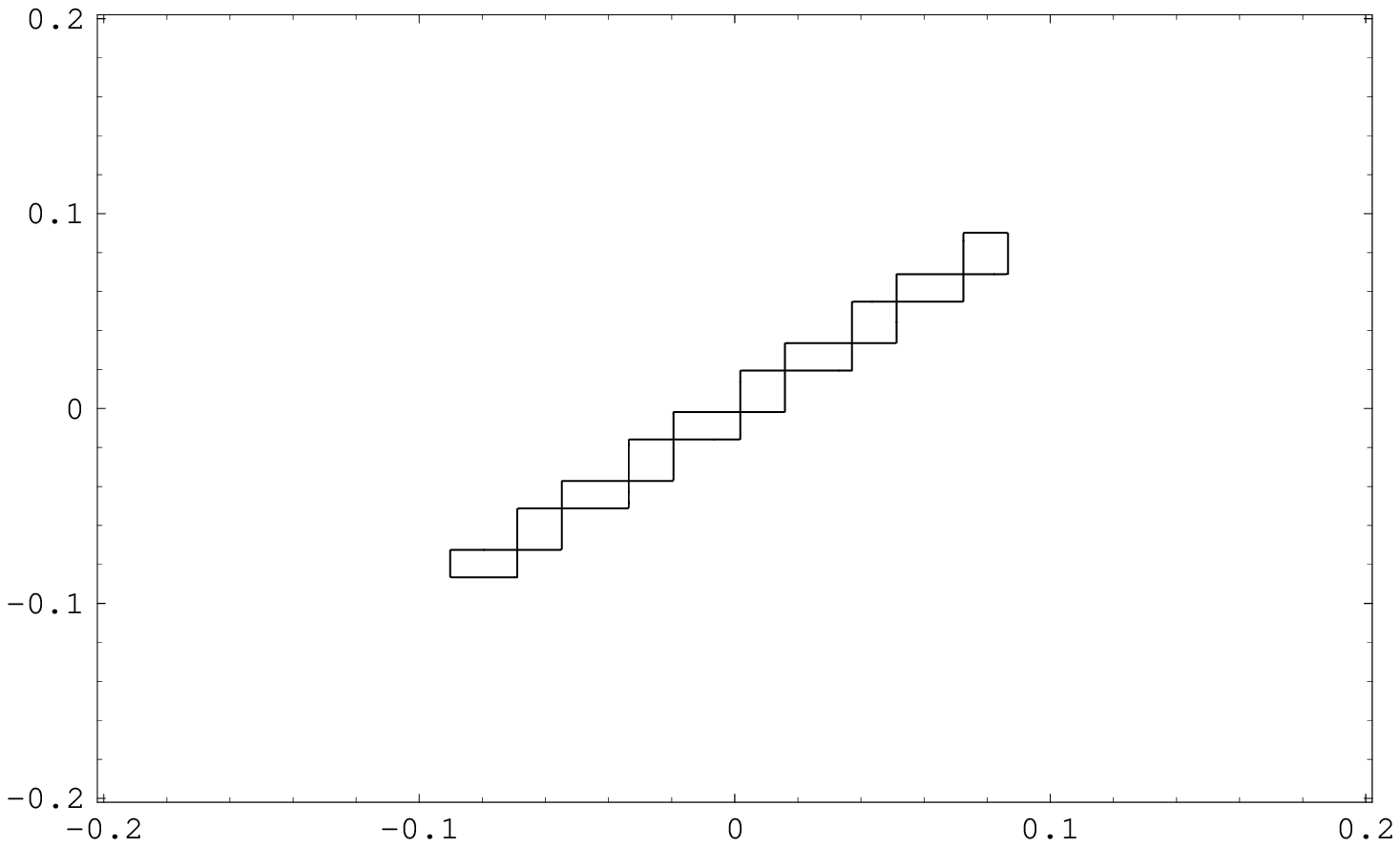,width=3.0in,angle=0}
\caption{The string solution for $n=10$, $q=1$, at $\tau=0$(left figure) and $\tau=0.035L$(right figure);
the axes(horizontal axis: $\phi$; vertical axis: $\gamma$) are labeled in units of $\kappa L$}
\label{figure_n10}
\end{figure}

\begin{figure}[ht]
\epsfig{figure=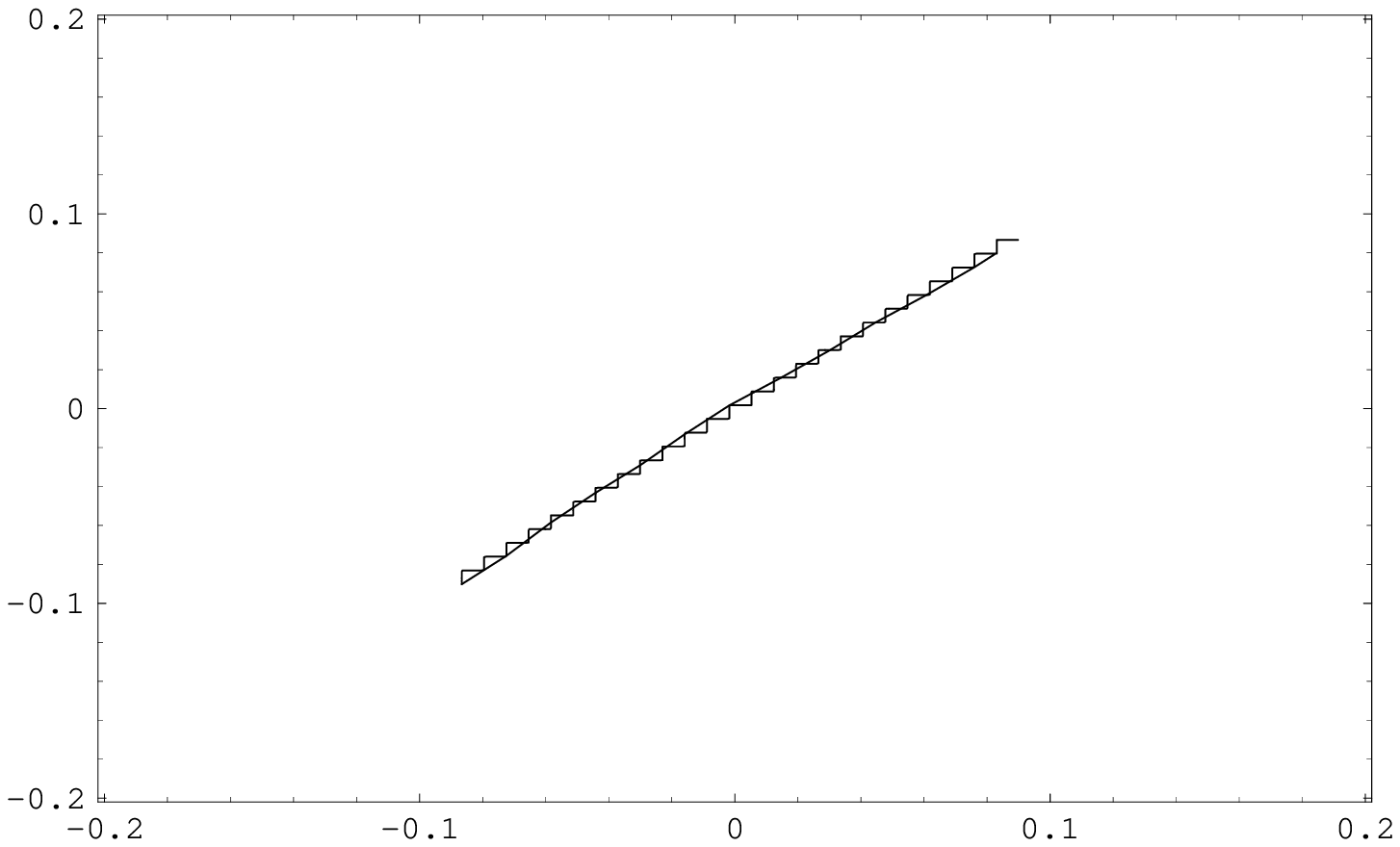,width=3.0in,angle=0}
\epsfig{figure=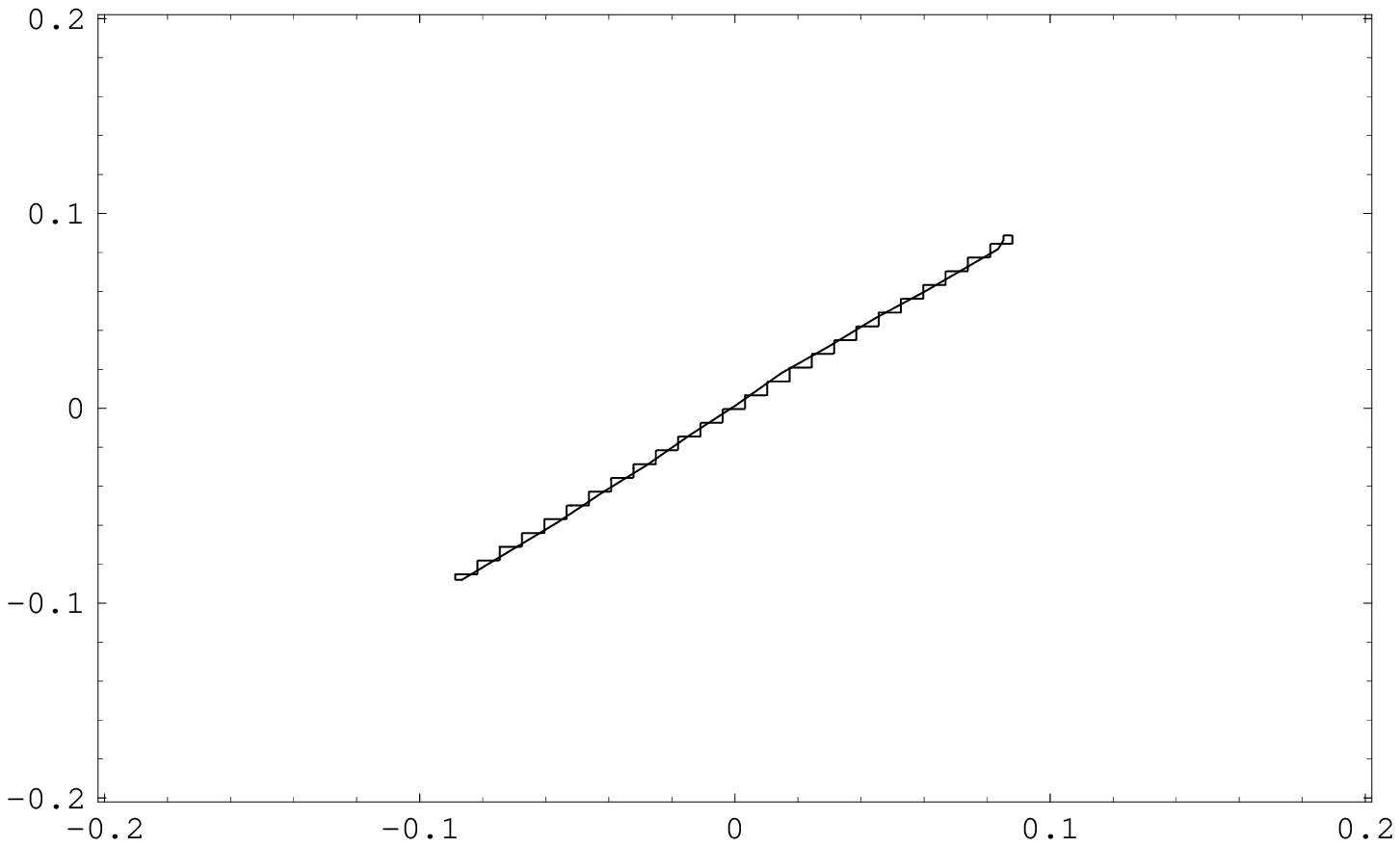,width=3.0in,angle=0}
\caption{The string solution for $n=50$, $q=1$, at $\tau=0$(left figure) and $\tau=0.003L$(right figure);
the axes(horizontal axis: $\phi$; vertical axis: $\gamma$) are labeled in units of $\kappa L$}
\label{figure_n50}
\end{figure}

What is quite interesting is the large $n$ limit. In this limit, $\gamma\approx F(\xi^+)$, $\phi\approx qF(\xi^+)$,
and the
string looks roughly like a static straight line extending from $\left(\gamma=-\frac{\kappa}{\sqrt{1+q^2}}\frac{L}{8},
\phi=-q\frac{\kappa}{\sqrt{1+q^2}}\frac{L}{8}\right)$ and $\left(\gamma=\frac{\kappa}{\sqrt{1+q^2}}\frac{L}{8},
\phi=q\frac{\kappa}{\sqrt{1+q^2}}\frac{L}{8}\right)$. However, when zooming in, one sees a lot of small rectangles oscillating
very rapidly (with period $\delta\tau\sim\frac{L}{2n}$), which come from the right moving part $\frac{1}{n}F(n\xi^-)$.
These fast oscillating tiny kinks account for the acceleration caused by the string's tension and prevent the string
from contracting. See Figure \ref{figure_n10} and Figure \ref{figure_n50}.

\section{Comparison with the ``2d-dual'' version of NR system}\label{section_NR_system}
In this section, we shall compare our solutions given in Section \ref{section_our_solution} with the ``2d-dual''
 of Neumann-Rosochatius(NR) system
and show their differences. As explained in \cite{Arutyunov:2003za,Tseytlin:2003ii}, interchanging the roles
of the worldsheet coordinates $\tau$ and $\sigma$ (the ``2d-dual'' version) turns rigidly rotating strings into
pulsating strings. Let us briefly review this material below.

The rotating string ansatz on $S^5$ is
\BE
X_1+iX_2=z_1(\sigma)e^{i\omega_1\tau},\ X_3+iX_4=z_2(\sigma)e^{i\omega_2\tau},\ X_5+iX_6=z_3(\sigma)e^{i\omega_3\tau}
\EE
with $z_i(\sigma)=r_i(\sigma)e^{i\alpha_i(\sigma)},i=1,2,3$. Interchanging $\tau\leftrightarrow\sigma$ we get the pulsating
string ansatz
\BE
X_1+iX_2=z_1(\tau)e^{im_1\sigma},\ X_3+iX_4=z_2(\tau)e^{im_2\sigma},\ X_5+iX_6=z_3(\tau)e^{im_3\sigma}
\EE
with $z_i(\tau)=r_i(\tau)e^{i\alpha_i(\tau)},i=1,2,3$. Periodicity of the $X$'s implies that the $m_i$'s must be
integers. Plugging this ansatz into eqn. (\ref{eqn_o6_model}), we find the unit modulus condition becomes
\BE
\sum_{i=1}^3 r_i^2=1
\EE
the conformal gauge conditions become
\BE
& &\sum_{i=1}^3\left(\dot{r}_i^2+r_i^2\dot{\alpha}_i^2+r_i^2m_i^2\right)=\kappa^2\nonumber\\
& &\sum_{i=1}^3 m_ir_i^2\dot{\alpha}_i=0
\EE
and the equations of motion become
\BE
& &\ddot{r}_i-r_i\dot{\alpha}_i^2+\left(m_i^2-\Lambda\right)r_i=0\nonumber\\
& &r_i\ddot{\alpha}_i+2\dot{r}_i\dot{\alpha}_i=0
\EE
with $-\Lambda=\sum_{i=1}^3\left(\dot{r}_i^2+r_i^2\dot{\alpha}_i^2-r_i^2m_i^2\right)$. Integrating the above equation of
motion for $\alpha_i$ once we get three constants of motion (which are angular momenta)
\BE
r_i^2\dot{\alpha}_i={\cal J}_i=\text{const}
\EE
The resulting class of pulsating strings are described by the effective particle-mechanics Lagrangian of the
NR integrable system and can be studied in details, for example, one can take the oscillation ``level number'' $N$ to be
large and consider the $\frac{\lambda}{N^2}$ expansion of the string's energy.

The above pulsating string ansatz only gives a subset of all the solutions on $S^5$, which is due to the fact that
it has a relatively simple $\sigma$-dependence ($e^{im_i\sigma}$) and its $\tau$-dependence and $\sigma$-dependence are
``separated''. In particular, the solutions constructed in Section \ref{section_our_solution} depend on both left
and right moving coordinates $\tau\pm\sigma$ in a crucial manner and hence are not included in the NR system pulsating string
ansatz. We now illustrate this point by focusing on solutions lying completely in the $X_1-X_2$ plane. For the NR system,
this corresponds to setting $r_2=0,r_3=0$, which leads to $r_1=1$. The conformal gauge conditions become
\BE
\dot{\alpha}_1^2+m_1^2=\kappa^2,\ \ m_1\dot{\alpha}_1=0
\EE
The equations of motion for the $\alpha_i$'s give
\BE
{\cal J}_1=\dot{\alpha}_1,\ \ {\cal J}_2=0,\ \ {\cal J}_3=0
\EE
while the equations of motion for the $r_i$'s are automatically satisfied. Hence there are only two possible solutions
depending on what value the angular momentum ${\cal J}_1$ takes. If we choose ${\cal J}_1\neq 0$, i.e. $\dot{\alpha}_1\neq 0$,
then the second equation
of the conformal gauge conditions gives $m_1=0$ and we get
\BE
X_1+iX_2=e^{i\alpha_1}=e^{i({\cal J}_1\tau+\alpha_1(0))}
\EE
which is a point-like string orbiting on the $S^1$ in the $X_1-X_2$ plane, with an energy (using the first equation
of the conformal gauge conditions) $\frac{E}{\sqrt{\lambda}}=\kappa=|{\cal J}_1|$. If we choose ${\cal J}_1=0$, i.e. $\dot{\alpha}_1=0$, then the first equation of
conformal gauge conditions gives $\kappa=|m_1|$ and we get
\BE
X_1+iX_2=e^{i\alpha(0)+m_1\sigma}
\EE
which is a static string wound $m_1$ times around the $S^1$ in the $X_1-X_2$ plane. Next let us look at the solution
given as Example {\bf I} of Section \ref{subsection_example_solutions} with $q=\infty,n=1$. This gives
$\gamma=0,\phi(\tau,\sigma)=\Phi(\xi^+)+\tilde{\Phi}(\xi^-)\neq 0$, which is a folded string
pulsating on the $\phi$-axis with its center at rest and its length (in terms of the range of $\phi$) changing periodically between $0$
and $\kappa\frac{L}{2}$ (see Section \ref{subsection_example_solutions}). Translating to the embedding Cartesian
coordinates this is
\BE
X_1+iX_2=e^{i\phi(\tau,\sigma)}, X_3+iX_4=0,\ \ X_5+iX_6=0
\EE
describing a folded string pulsating on the $S^1$ in the $X_1-X_2$ plane with its length changing periodically. This solution
is obviously distinct from the point-like string and the static wound string obtained from the NR system.

\section {Fluctuation Spectrum}\label{section_fluctuation}
Given a classical solution of the string equations of motion it is naturally of interest
to study the fluctuations in this background and for other semi-classical strings this has proved
possible in some cases e.g. \cite{Frolov:2002av,Frolov:2003tu}.
It is possible to consider fluctuations about the most general kinky string solution 
that we found above, however the formula are complicated and the resulting fluctuation 
action has non-constant coefficients. We will restrict our attention to the simple 
case of the pulsating folded string and in particular the $q\rightarrow \infty$, $n=1$ case (Example ${\bf I}$ in 
Section \ref{subsection_example_solutions}). 
Starting from the
Lagrangian
\BE
{\cal L}=-\frac{\sqrt{\lambda}}{4 \pi}\ G_{\mu \nu} \partial_{\alpha}X^{\mu}\partial^{\alpha} X^{\nu}
\EE
in conformal gauge and with the metric (here let us consider the only the time and sphere part)
\BE
ds^2= -dt^2+d\gamma^2+\cos^2\gamma\ d\phi_1^2+\sin^2\gamma\  \left(d\psi^2+\cos^2\psi\ d\phi_2^2+\sin^2 \psi\ d\phi_3^2\right).
\EE
We define the fields $\phi=\frac{1}{2}(\phi_1+\phi_2)$,  ${\hat \phi}=\frac{1}{2}(\phi_1-\phi_2)$ and expand 
$t=\kappa \tau+{\tilde t}$, $\phi=\phi_0+{\tilde \phi}$, ${\hat \phi}={\tilde {\hat \phi}}$ and $\gamma={\tilde \gamma}$. It does not matter what values $\psi$ and $\phi_3$ take in the classical solution so  we do not expand these fields. We get
\BE
G_{\mu\nu} \partial_{\alpha}X^{\mu}\partial^{\alpha} X^{\nu}&=& \kappa^2 +2\kappa \partial_{\tau}{\tilde t}+(\partial {\tilde \phi})^2+(\partial {\tilde \gamma})^2
+\left(1-\frac{{\tilde \gamma}^2}{2}\right)^2(\partial \phi_0+{\tilde \phi}+{\tilde {\hat \phi}})^2\nn\\
& &+{\tilde \gamma}^2 ((\partial \psi)^2+\sin^2 \psi\partial \phi_3^2)+\tilde{\gamma}^2\cos^2 \psi (\partial \phi_0)^2.
\EE
where $(\partial A)^2=-(\partial_{\tau} A)^2+(\partial_{\sigma} A)^2$. We now write the fluctuations of the spherical coordinates in terms of Cartesian variables $x^i$ 
\BE 
(\partial {\tilde \gamma})^2 +{\tilde \gamma}^2(\partial \psi^2 +\sin^2 \psi \partial \phi_3^2)=\sum_{i=1}^3(\partial x^i)^2\nn\\
{\tilde \gamma}^2=\sum_{i=1}^3(x^i)^2\EE
and so quadratic fluctuation action is 
\BE
{\cal L}\sim -(\partial{\tilde  t})^2 +(\partial {\tilde \phi})^2+(\partial {\tilde {\hat \phi}})+\sum_{i=1}^3 (\partial x^i)^2+(-1+\cos^2\psi)(\partial \phi_0)^2 (x^i)^2.
\EE

However for the simple folded string we do not need to introduce the new coordinate $\phi$ but rather we can simply expand about the solution 
$t=\kappa\tau+{\tilde t}$, $\phi_1=\phi_0+{\tilde \phi_1}$ and $\gamma=0$ with 
the rest of the coordinates being arbitrary. We thus find for the quadratic 
fluctuations
\BE
{\cal L} \sim -(\partial {\tilde t})^2+(\partial{ \tilde \gamma})^2 
+\left(1-\frac{{\tilde \gamma}^2}{2}\right)^2(\partial \phi_0+\partial {\tilde \phi_1})^2
+{\tilde \gamma}^2(\partial \Omega_3)^2.
\EE
Where $\Omega_3$ represents the angles on a three sphere and as before we introduce Cartesian 
variable  $(\partial {\tilde \gamma})^2+{\tilde \gamma}^2(\partial \Omega_3)^2=\sum_{i=1}^{4}(\partial x^i)^2$. We now have
\BE
{\cal L}\sim -(\partial {\tilde t})^2+(\partial {\tilde \phi_1})^2+\sum_{i=1}^4 (\partial x^i)^2-(\partial \phi_0)^2 (x^i)^2.
\EE
Explicitly, we expand about the solution $\gamma=0$ and $\phi_1=qF(\xi^+)-qF(\xi^-)$ 
and $F(\xi)=\int_0^{\xi} f({\tilde \xi})d{\tilde \xi}$ with the $f$'s as in (\ref{feqn}) and we take $q\rightarrow \infty$. Our solution breaks into different regions given by
\BE
\begin{array}{lcr}
\mbox{\bf Region I}  & 0\leq \xi^+ \leq \frac{L}{4}            & \\
                     & 0\leq \xi^- \leq \frac{L}{4}            & \phi_1 =  \kappa \sigma \\
& & \\
\mbox{\bf Region II} & 0\leq \xi^+ \leq \frac{L}{4}            & \\
                     & \frac{L}{4} \leq \xi^- \leq \frac{L}{2} & \phi_1 = \kappa (\tau -L/2)\\
& & \\
\mbox{\bf Region III}& \frac{L}{4}\leq \xi^+ \leq \frac{L}{2}  & \\
                     & 0\leq \xi^- \leq \frac{L}{4}            & \phi_1 = -\kappa (\tau -L/2)\\
& & \\
\mbox{\bf Region IV} & \frac{L}{4}\leq \xi^+ \leq \frac{L}{2}  & \\
                     & \frac{L}{4} \leq \xi^- \leq \frac{L}{2} & \phi_1 = -\kappa \sigma.\\
\end{array}
\EE
Thus the bosonic fluctuations of the $S^5$ coordinates in region II are described by,
\BE
L&=&-\frac{1}{4 \pi} \Big( \sum_{i=1}^{4}(-\pt x_i^2+\ps x_i^2+\kappa^2 x_i^2)-\kappa^2
-2 \lambda^{\frac{1}{4}} \kappa \pt {\tilde  \phi_1} - \pt{\tilde \phi_1}^2+\ps{\tilde \phi_1}^2 \Big)
\EE
where we see that the transverse coordinates have a positive mass term just as is found in the BMN 
limit, moreover the fluctuations in region III are the same except that the sign of the term linear 
in $\partial_{\tau} \phi_1$ is positive. However in regions I and IV we find the the mass term for the  transverse 
coordinates is negative. For example in region I
\BE
L&=&-\frac{1}{4 \pi} \Big( \sum_{i=1}^{4}(-\pt x_i^2+\ps x_i^2-\kappa^2 x_i^2)+\kappa^2
+2 \lambda^{\frac{1}{4}}\kappa \ps {\tilde \phi_1}-\pt {\tilde \phi_1}^2+\ps {\tilde \phi_1}^2\Big)
\EE
with region IV having a different sign on the term linear in $\ps \phi_1$. The fluctuations on 
the $AdS_5$ are straightforward
\BE
L_{\rm{AdS}}&=& -\frac{1}{4 \pi}\Big(\kappa^2+2\kappa \pt {\tilde t}+ \pt {\tilde t}^2 -\ps {\tilde t}^2+\sum_{i=1}^{4}(-\pt z_i^2+\ps z_i^2+\kappa^2 z_i^2)\Big)
\EE
where we here use the Cartesian coordinates made from fluctuations in the radial direction $\rho$ 
and the angular coordinates of the $S^3$ inside $AdS_5$, here the mass term is of course positive. 
We would like to remove the longitudinal fluctuations and to this end we consider the Virasoro constraints 
to linear order in fluctuations where we once again find different relations in different regions
\BE
\begin{array}{lcr}
\mbox{\bf Region I}  & \pt {\tilde t}=\ps {\tilde \phi},\   \ps {\tilde t}=\pt {\tilde \phi}  & \\
\\
\mbox{\bf Region II} &  \pt {\tilde t}=\pt {\tilde \phi},\   \ps {\tilde t}=\ps {\tilde \phi} & \\
\\
\mbox{\bf Region III}&  \pt {\tilde t}=-\pt {\tilde \phi},\   \ps {\tilde t}=-\ps {\tilde \phi}  & \\
\\
\mbox{\bf Region IV} & \pt {\tilde t}=-\ps {\tilde \phi},\  \ps {\tilde t}=-\pt {\tilde \phi}.  & \\
\\
                    \\
\end{array}
\EE
Thus the fluctuations in the $\phi$ direction can be related to those in the time direction which 
we can remove by rescaling the worldsheet time coordinate such that $t=\kappa \tau$. Our final result 
is that we find a simple  theory of transverse bosons in a quadratic potential. This potential corresponds to a simple real mass term in certain regions of the worldsheet though in others it 
corresponds to an imaginary mass. 
This imaginary mass is suggestive of the fact that our solution 
may not be stable. We should now proceed to solve the equations of motion for these 
varying quadratic potentials and explicitly check for tachyonic modes, however this is a 
little complicated so we will postpone it for future work and instead 
consider some of the possibilities for string decay.

\section{Decay of the string}\label{section_decay_of_strings}
The decay of highly excited strings in flat space has been under
investigation for some time, recent works being, for example,
\cite{Iengo:2002tf,Iengo:2003ct,Chialva:2003hg,Chialva:2004xm,
Iengo:2006gm} . Let us first review these flat space results to gain
some intuition. In \cite{Iengo:2002tf} very massive closed strings
which are maximum angular momentum eigenstates were considered, and
its quantum decay rate was obtained from a one-loop calculation. It
was found that the dominant decay channel is when $M_1,M_2$ (masses
of the two outgoing states) are both of the same order as $M$ (mass
of the initial string) and are restricted along a curve
$M_2=F(M_1)$, with the decay rate given by $\Gamma(M_1,M_2)\sim
M^{-4}$. The sub-dominant channel is when one of the outgoing states
is massless and the other one is very massive, with $\Gamma(M_1\sim
M,M_2=0)\sim M^{-5}$. The decay rates of all other channels, such as
$(M_1,M_2)$ being off the curve $M_2=F(M_1)$, or the two outgoing
states both being massless, are exponentially suppressed. When all
the decay channels were summed up, the lifetime of initial string
state was found to be ${\cal T}\sim M$. The initial highly excited
string with maximum angular momentum can be represented as a
spinning folded string, and a semi-classical analysis was carried
out in \cite{Iengo:2003ct} for the splitting of such a string into
two strings. The masses $M_1,M_2$ of the outgoing strings were
expressed as functions of $\sigma_b$, the point at which the initial
string splits, and then $\sigma_b$ was eliminated to give a relation
$M_2=\tilde{F}(M_1)$. Amazingly, it was found that the quantum and
semi-classical analysis give the same relation, i.e., $F$ and
$\tilde{F}$ are identical.

In \cite{Chialva:2003hg}, decay in flat space of very massive string states which are not maximum angular momentum
eigenstates was considered. In general, these strings are
not folded: they are rotating ellipses and their deformations. Since for these strings there is never a contact
between two different points of the string, they will not break
classically. Indeed, the one-loop computation of \cite{Chialva:2003hg} showed that the dominant decay channel is the emitting of massless particles,
i.e. one outgoing state being massless while the other outgoing state being massive. All other channels, such as string
splitting where both outgoing states are massive, are exponentially suppressed. Among these string states, the special case
of a rotating circular string was found to be most stable, with a lifetime ${\cal T}\sim M^5$. In \cite{Chialva:2004xm}
it was shown that the decay of this rotating circular string can be described by regarding it as a classical radiating
antenna. The more recent work \cite{Iengo:2006gm} constructed flat space string solutions which are rotating and pulsating
ellipses that become folded at an instant of time. These strings can break at the moment they become folded.
\cite{Iengo:2006gm} showed that the quantum decay rate of the massive channel for these strings agrees with a
semi-classical analysis of their splitting. Other flat space string configurations, such as those having two points that
get in contact at an instant of time, and their breaking were also investigated in \cite{Iengo:2006gm}.

It is natural to consider the generalization from flat space to $AdS_5\times S^5$. For example, a similar analysis of
the splitting of the folded two-spin string solution of \cite{Frolov:2003xy} was carried out in \cite{Peeters:2004pt}.
We shall consider the two-parameter $(q,n)$ family of strings given in Section \ref{subsection_example_solutions},
which is expected to have various breaking patterns similar to those in flat space.
When $q=0$ or $q=\infty$ the string is folded all the time and can break at any moment. When $q$ is nonzero and finite the string is
folded at an instant of time. Furthermore, when $n>1$ the string becomes self-intersecting, for example, as depicted in
the second plot of Figure \ref{figure_selfcrossing_n2} for the $n=2$ case. Below let us first consider the splitting of
the string using a semi-classical analysis. As in flat space this should shed light on the massive decay channels of
the quantum process.

\subsection{Splitting of the $n=1$ string solutions}
\begin{figure}[ht]
\epsfig{figure=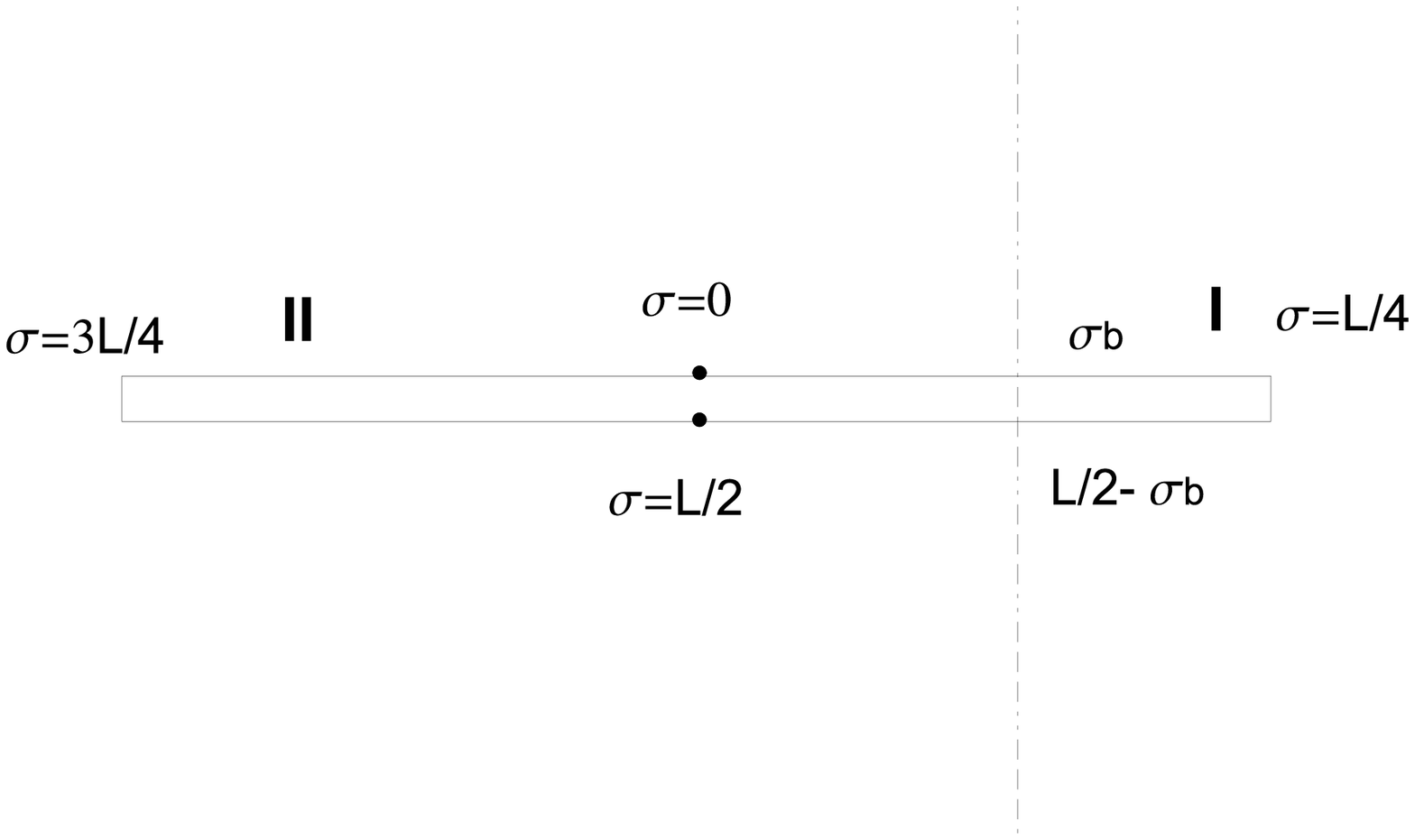,width=3.0in,angle=0}
\caption{splitting the string ($n=1$, $q=\infty$)}
\label{figure_splitn1qinfinite}
\end{figure}

At any given time, the string can spontaneously
split if two points on the string labeled by $\sigma$ and $\tilde{\sigma}$ have the same locations in the
target space, namely $\gamma(\tau,\sigma)=\gamma(\tau,\tilde{\sigma}), \phi(\tau,\sigma)=\phi(\tau,\tilde{\sigma})$.
Let us first look at the $q=\infty$ case where the string is folded along the $\phi$-axis. Using the fact that
$f(\xi+\frac{L}{4})=-f(\xi)$ and $F(\xi+\frac{L}{4})=-F(\xi)$ one sees that at any given time, the points labeled
by $\sigma$ and $\frac{L}{2}-\sigma$ have the same $\phi$ (and also the same $\dot{\phi}$). Hence let's suppose the string splits
at a point labeled by $\sigma_b$ (and $\frac{L}{2}-\sigma_b$) at time $\tau_b$. See Figure \ref{figure_splitn1qinfinite}.
 We compute the charges of the two
outgoing pieces I and II by evaluating them at time $\tau_b$ just before the string splits.
The energies are
\BE
E^I=\sqrt{\lambda}\int_{\sigma_b}^{\frac{L}{2}-\sigma_b}\frac{d\sigma}{L}\kappa=\sqrt{\lambda}\ \kappa
\left(\frac{1}{2}-\frac{2\sigma_b}{L}\right),\ \ E^{II}=\sqrt{\lambda}\ \kappa
\left(\frac{1}{2}+\frac{2\sigma_b}{L}\right)
\EE
The angular momenta are
\BE
J^I_{12}=\sqrt{\lambda}\int_{\sigma_b}^{\frac{L}{2}-\sigma_b}\frac{d\sigma}{L}\frac{1}{2}(\phi_+ +\phi_-)
=-2\frac{\sqrt{\lambda}}{L}q\left[F(\frac{\tau_b+\sigma_b}{2})+F(\frac{\tau_b-\sigma_b}{2})\right]
\EE (where it is understood that the $q\to\infty$ limit is taken)
and $J^{II}_{12}=-J^I_{12}$. All other components $J^{I,II}_{AB}=0$. There is a clear interpretation of this:
before splitting, the string pulsates along the $\phi$ direction, namely, it lies on a circle in the $X_1-X_2$ plane
with its center of mass at rest and its length changing periodically; after the splitting, one piece goes counter-clockwise
and the other goes clockwise on the circle. One remark is in order. $J^I_{12}$ depends on both when ($\tau_b$) and where
($\sigma_b$) the string splits, hence
one cannot get a energy-angular momentum relation $E^I(J^I_{12})$ by simply eliminating $\sigma_b$. Put another way, the outgoing
strings have a continuous spectrum of energy-angular momentum relation parametrized by $\tau_b$. This is in contrast to
the case considered in \cite{Peeters:2004pt}, where the folded string is a rotating rigid body and hence the energy
and angular momenta of the outgoing pieces only depend on where the breaking point is.

The $q=0$ case is similar. Now the string is folded along the $\gamma$-axis. And at any given time
points labeled by $\sigma$ and $L-\sigma$
have the same $\gamma$ and $\dot{\gamma}$. Hence we consider the string splitting at the point labeled by
$\sigma_b$ (also $L-\sigma_b$) at time $\tau_b$. We find
\BE
E^I=\sqrt{\lambda}\int_{-\sigma_b}^{\sigma_b}\frac{d\sigma}{L}\kappa=\sqrt{\lambda}\ \kappa\frac{2\sigma_b}{L},\
\ E^{II}=\sqrt{\lambda}\ \kappa\left(1-\frac{2\sigma_b}{L}\right)
\EE
Also
\BE
J^I_{13}=2\frac{\sqrt{\lambda}}{L}\left[F(\frac{\tau_b+\sigma_b}{2})-F(\frac{\tau_b-\sigma_b}{2})\right],\ \
J^{II}_{13}=-J^I_{13}
\EE
while all other components $J^{I,II}_{AB}$ vanish. The interpretation is also similar: before the splitting, the string
lies on a circle in the $X_1-X_3$ plane with its center of mass at rest and its length changing periodically; after the
splitting the two outgoing strings go counter-clockwise and clockwise respectively.

For finite nonzero values of $q$ the string is the pulsating rectangle, which at $\tau=0$ gets folded along the $\gamma$-axis
, at $\tau=\frac{L}{4}$ gets folded along the $\phi$-axis, and so forth. At $\tau=0$, points at $\sigma$ and $L-\sigma$
have the same $\gamma$, and the same $\dot{\gamma}$ (as a matter of fact, $\dot{\gamma}(\tau=0,\sigma)=0$ throughout the
string). Their velocities in the $\phi$-direction are nevertheless opposite to each other: $\dot{\phi}(\tau=0,L-\sigma)=-\dot{\phi}(\tau=0
,\sigma)=-qf(\frac{\sigma}{2})$. This is because at $\tau=0$ the string is starting to re-expand in the $\phi$-direction
(see Figure \ref{figure_n1}). The energies of the two outgoing pieces are:
\BE
E^I=\sqrt{\lambda}\ \kappa\frac{2\sigma_b}{L},\
\ E^{II}=\sqrt{\lambda}\ \kappa\left(1-\frac{2\sigma_b}{L}\right)
\EE
while all the angular momenta vanish
\BE
J^I_{AB}=0,\ \ J^{II}_{AB}=0
\EE
since for finite nonzero $q$, all of the angular momentum densities are
total derivatives of functions of either $\gamma$ or $\phi$ w.r.t to $\sigma$ (see eqn.~(\ref{eqn_J_density})). At $\tau
=\frac{L}{4}$ similarly one finds
\BE
E^I=\sqrt{\lambda}\ \kappa
\left(\frac{1}{2}-\frac{2\sigma_b}{L}\right),\ \ E^{II}=\sqrt{\lambda}\ \kappa
\left(\frac{1}{2}+\frac{2\sigma_b}{L}\right)
\EE
and
\BE
J^I_{AB}=0,\ \ J^{II}_{AB}=0
\EE
\subsection{Splitting of the $n>1$ string solutions}
The $n>1$ case bears a strong resemblance to the $n=1$ case. More specifically, when $q=\infty$ or $q=0$ the string is folded
all the time and can break at any time into two pieces each carrying non-vanishing angular momentum $J_{12}$ or $J_{13}$;
 when $q$ is nonzero and finite, the string becomes folded at some instant of time and can break at that time into two
 pieces both carrying zero angular momentum. However, one new feature that arises in the $n>1$ case, for nonzero and finite
 $q$, is its self-crossing and breaking at the self-crossing points.

Let us begin with the $n=2$ case (see Figure \ref{figure_selfcrossing_n2}). At a generic time $\tau$ it has a figure ``$8$''
shape.
\begin{figure}[ht]
\epsfig{figure=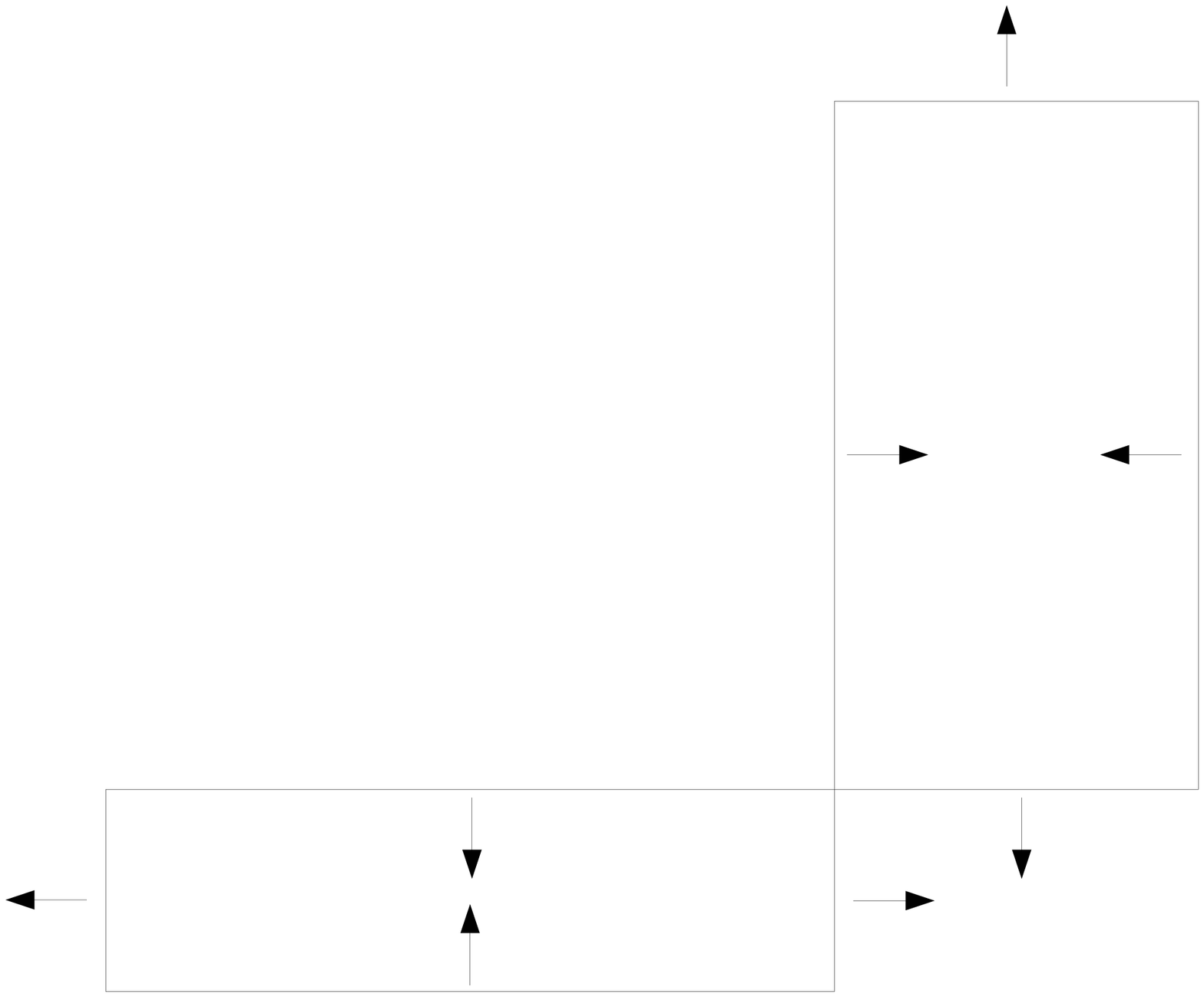,width=3.0in,angle=0}
\caption{The string solution for $n=2$ at a generic time $\tau$; velocities of the segments are indicated by arrows;
horizontal axis: $\phi$; vertical axis: $\gamma$}
\label{figure_n2_velocity}
\end{figure}
The self-crossing point is parametrized by $\sigma=\frac{L}{4}-\tau$ and $\tilde{\sigma}=\frac{3L}{4}-\tau$. Breaking
at the self-crossing point gives two outgoing strings with equal energies $E^I=E^{II}=\frac{\sqrt{\lambda}}{2}\kappa$, and
vanishing angular momenta $J^{I}_{AB}=J^{II}_{AB}=0$. We have plotted in Figure \ref{figure_n2_velocity} the velocities
of the different segments of the string before the breaking. From this, we see that the outgoing strings are two $n=1$
strings at different stages of their individual evolution.

Other $n>1$ solutions have similar behavior. At a generic time, it has $n-1$ nodes (see Figure \ref{figure_n10} for the
$n=10$ example), and can break at any of these nodes into an $(n_1=n-1)$-string and an $(n_2=1)$-string. The $(n-1)$-string
can undergo further splitting in a similar fashion. In this sense, we can say that the $(n=1)$-string is the building block of
all the $n>1$ strings.

\section{Discussion}\label{section_discussion}

As mentioned previously, kinky strings have been studied in the context of
cosmic strings in a flat target space. Additionally, strings with discontinuous worldsheet first derivatives 
have been studied in some detail in both flat and curved two-dimensional target spacetimes 
(e.g. \cite{Patrascioiu:1974un},\cite{Bardeen:1975gx, Bardeen:1976yt}, 
\cite{Bars:1993sq}, \cite{Bars:1994xi}). It was shown that longitudinal 
kink modes can be found by carefully taking the massless limit 
of the massive Nambu string and in $2d$ flat spacetime, it was
shown that the longitudinal kink modes of the folded string could be 
recast as integrable system of 
point-like particles connected by linear potentials; the
action-angle variables and subsequent canonical quantization of
these point-like degrees of freedom were also investigated. These kinky strings 
may be related to some version of QCD with matter in the adjoint. Interestingly
these folded strings have recently appeared in the context of the $AdS/CFT$ 
correspondence \cite{Klebanov:2006jj} where they were considered as the dual description of 
the excitations of flux tubes connecting quark/anti-quark pairs.  
The kinky solutions we consider in this paper
are parametrized by a constant $q$, which in the $q\to 0,\infty$
limit become folded strings on a $S^1\subset S^5$ and their kinks
are exactly analogues of the longitudinal modes of $2d$ strings in
flat spacetime. For generic values of $q$ our solutions are no
longer folded and have transverse modes (they move on an $S^3\subset
S^5$), thus being more complicated. The construction of action-angle
variables for the kinks of these solutions and their canonical
quantization definitely deserve more investigation as do their relationship
to excitations in the gauge theory.

The kinky solutions considered in this article all have vanishing angular momenta, i.e. 
the string's center of mass is at rest on the $S^5$. One should be able to boost these solutions
to give them angular momenta by relaxing the condition that  
$\Gamma,\tilde{\Gamma},\Phi,\tilde{\Phi}$ have to be periodic in their arguments $\xi^+$ and 
$\xi^-$ (see
the paragraph under eqn. (\ref{eqn_soln3})). For example, we can boost the $q=\infty$ folded  
string solution $\phi(\tau,\sigma)$ by shifting it to $\tilde{\phi}(\tau,\sigma)=\phi(\sigma,\tau)+\omega\tau$, which 
corresponds to giving linear shifts to the left-moving and right-moving parts $\Phi(\xi^+)\to\Phi(\xi^+)+\omega\xi^+$,  
$\tilde{\Phi}(\xi^-)\to\tilde{\Phi}(\xi^-)+\omega\xi^-$ thus breaking their periodicity (the string embedding
$\tilde{\phi}(\tau,\sigma)$ is of course still periodic under $\sigma\to\sigma+L$). 
One can similarly boost the generic kinky solutions and  then  take the  angular momentum $J$ to be large. 
We would like to emphasize that our kinky solutions are different from the ``giant magnon''
solutions recently constructed in \cite{Hofman:2006xt} in that the kinky strings are non-rigid pulsating while 
the giant magnons are rigid strings; furthermore
the kinky strings are on $R_t\times S^3$ while the giant magnons are on $R_t\times S^2$.
 They do share the common feature of being made
of straight string segments (although, the kinky strings are in the standard spherical coordinates while 
the giant magnons are made of straight lines only when cast 
in LLM coordinates). In fact the giant magnons seem to be closely related 
to the spiky strings considered in \cite{Kruczenski:2004wg} and in particular to those in \cite{Ryang:2005yd}
where the solution of Kruczenski was generalized to a two sphere inside $S^5$.

As it turns out, there are more kinky solutions in flat space, which we have not yet
been able to generalize to $R_t\times S^5$. For example, there is a family of solutions in flat space which
we call ``pulsating polygonal strings'', describing an $N$-sided polygonal string pulsating in 
the $(X_1,X_2)$ plane. This solution is obtained by setting
\BE \label{eqn_pulsating_polygon_ab}{\bf a}^{(1)}(u)={\bf f}(u),\ \ {\bf b}^{(1)}(v)={\bf f}(v)\EE
with the function ${\bf f}(z)$ given by, written as a complex number $f(z)$
\BE\label{eqn_pulsating_polygon_f} f(z)=\exp\left[i\frac{2\pi}{N}int\left(\frac{Nz}{L}\right)\right]\EE
where $int(y)$ is the integer part of $y$, e.g. $int(0.65)=1$, $int(-0.3)=-1$.

For even values of $N$, it is easy to show that at $\tau=\frac{L}{4}$, the string shrinks to a point (its center
of mass). For odd values of $N$, the string never shrinks to a point. For even $N$ the evolution of the string's
profile is: the string starts out as an $N$-sided polygon at $\tau=0$; shrinks to a point at
$\tau=\frac{L}{4}$; expands back to the original polygon at $\tau=\frac{L}{2}$; so on and so forth.
% See Figure \ref{pulsating} for the $N=4$ case.
In the limit of $N\to\infty$, this solution reduces to the known pulsating circular string with its center at $(X_1=0,X_2=\frac{L}{2\pi})$.

Can one generalize this family of solutions to strings on $R_t\times S^5$? We can see that the approach in 
Section \ref{section_our_solution} does not generate this desired pulsating polygon
solution since the angles between the two string segments joining at the kinks are always $90$ degrees in all the solutions
obtained there. Let us see whether the NR system discussed in Section \ref{section_NR_system} can do the job. Let us 
first review a few pulsating
solutions that can be obtained from the NR system.
One example is the circular string pulsating on an $S^2\subset S^5$ \cite{Minahan:2002rc}\footnote{Similar pulsating circular string solutions were considered in \cite{Alishahiha:2002fi}}.
 To get this solution, one
takes $r_3=0$, $m_2=0,\alpha_2=0$ (see Section \ref{section_NR_system}), so that the string is moving on the $S^2$ embedded in $(X_1,X_2,X_3)$. One also takes
${\cal J}_1={\cal J}_2={\cal J}_3=0$ so that the string does not have any angular momentum. One also has to take $\alpha_1=0$
so that the $\alpha_1$ equation of motion is satisfied. Then as can be easily verified, this gives the circular pulsating
string, with
\BE
X_1+iX_2=r_1(\tau)e^{im_1\sigma},\ X_3=r_2(\tau), \ X_4=X_5=X_6=0
\EE
with $r_2=\sqrt{1-r_1^2}$ and $r_1(\tau)$ satisfying
\BE
\ddot{r}_1+\left[\frac{\dot{r}_1^2}{1-r_1^2}+m_1^2\left(1-r_1^2\right)\right]r_1=0
\EE
and the energy being
\BE
\frac{E}{\sqrt{\lambda}}=\kappa=\sqrt{m_1^2r_1^2+\frac{\dot{r}_1^2}{1-r_1^2}}
\EE
In the special case of $m_1=0$, the above solution reduces to a point like string orbiting in the $X_1-X_3$ plane, with
$X_1=\cos(\omega\tau+\theta_0)$, $X_2=0$, $X_3=\sin(\omega\tau+\theta_0)$, and $\frac{E}{\sqrt{\lambda}}=|\omega|$.
More pulsating string solutions from the NR system have also been obtained in \cite{Dimov:2004xi} for the $m_2\neq 0$ case
and ${\cal J}_1\neq 0, {\cal J}_2\neq 0$ case. However, a little thought reveals
that the NR system will not give a pulsating polygonal string, because it assumes the $r_i$'s to depend only on $\tau$
 but not on $\sigma$.

One natural question is what are the dual
gauge theory operators of these kinky solutions. As usual one would like to identify 
a large charge the inverse of which could be used for a power series expansion. Since the kinky solutions have vanishing angular 
momenta, one should identify a large oscillation quantum number in a way similar to the case of circular pulsating 
strings \cite{Minahan:2002rc}, which would require a better understanding of the point-like nature of the kink degrees of 
freedom, preferably at the level of the understanding in the $2d$ string context described at the beginning of this section. The dual gauge theory operators
should be easier to investigate after the solutions are boosted and the large angular momentum limit is then taken. 
We hope to have something more concrete to say regarding these issues in a future work.

\section*{Acknowledgements}
We would like to thank S. Das, J. Michelson, A. Mikhailov, R. Roiban, I. Swanson, and 
A. Tseytlin, for many useful comments and discussions. The work of X.W. is 
supported in part by Department of Energy
contract \#DE-FG01-00ER45832 and the National Science Foundation
grant No.~PHY-0244811. The work of TMcL is supported by funds from PSU.

%\begin{figure}[ht]
%\epsfig{figure=pulsating.eps,width=7.0in,angle=0}
%\caption{the solution given by eqn.'s~(\ref{eqn_pulsating_polygon_ab},\ref{eqn_pulsating_polygon_f}) for $N=4$.}
%\label{pulsating}
%\end{figure}

\end{document}